\documentclass[%
 aip,
 jmp,%
 amsmath,amssymb,
 reprint,%
]{revtex4-2}

\usepackage{graphicx}
\usepackage{dcolumn}
\usepackage{bm}
\usepackage{booktabs}
\usepackage{caption}
\usepackage{subcaption}
\usepackage{xcolor}
\usepackage[english]{babel}
\usepackage{soul}

\usepackage{hyperref}
\hypersetup{
    colorlinks=true,
    citecolor=blue,
    linkcolor=blue,
    filecolor=blue,      
    urlcolor=blue,
    pdftitle={Overleaf Example},
    pdfpagemode=FullScreen,
    }
\begin{document}

\preprint{AIP/123-QED}

\title{Accelerating Quantum Optimal Control of Multi-Qubit Systems with Symmetry-Based Hamiltonian Transformations}
\thanks{This article may be downloaded for personal use only. Any other use requires prior permission of the author and AIP Publishing. This article appeared in \emph{AVS Quantum Science} and may be found at \url{https://doi.org/10.1116/5.0162455}.}

\author{Xian Wang}
 \email{xwang056@ucr.edu}
 \affiliation{Department of Physics \& Astronomy, University of California-Riverside, 900 University Ave, Riverside, CA, 92521, United States}
\author{Mahmut Sait Okyay}%
\author{Anshuman Kumar}%
\affiliation{Materials Science \& Engineering Program, University of California-Riverside, 900 University Ave, Riverside, CA, 92521, United States}%

\author{Bryan M. Wong}
 \email{bryan.wong@ucr.edu}
 \affiliation{Department of Physics \& Astronomy, University of California-Riverside, 900 University Ave, Riverside, CA, 92521, United States}
 \affiliation{Materials Science \& Engineering Program, University of California-Riverside, 900 University Ave, Riverside, CA, 92521, United States}
 \affiliation{Department of Chemistry, University of California-Riverside, 900 University Ave, Riverside, CA, 92521, United States}

\date{\today}

\begin{abstract}
We present a novel, computationally efficient approach to accelerate quantum optimal control calculations of large multi-qubit systems used in a variety of quantum computing applications. By leveraging the intrinsic symmetry of finite groups, the Hilbert space can be decomposed and the Hamiltonians block diagonalized to enable extremely fast quantum optimal control calculations. Our approach reduces the Hamiltonian size of an $n$-qubit system from {$2^n \times 2^n$} to {$O(n \times n)$} or {$O(\frac{2^n}{n} \times \frac{2^n}{n})$} under $S_n$ or $D_n$ symmetry, respectively. Most importantly, this approach reduces the computational runtime of qubit optimal control calculations by orders of magnitude while maintaining the same accuracy as the conventional method. As prospective applications, we show that (1) symmetry-protected subspaces can be potential platforms for quantum error suppression and simulation of other quantum Hamiltonians, and (2) Lie-Trotter-Suzuki decomposition approaches can generalize our method to a general variety of multi-qubit systems.
\end{abstract}

\maketitle

\section{Introduction} \label{introduction}

The accurate and efficient control of qubit-based systems continues to attract significant interest due to their potential applications in high-performance algorithms \cite{shor1994,grover1996}, cryptography \cite{BB84, E91}, and quantum simulations \cite{britton2012}. In particular, manipulating qubits with tailored external pulses is one promising approach for realizing physical quantum gates for quantum information processing. In general, the field of quantum optimal control (QOC) focuses on constructing the temporal form of an optimized pulse to drive a system's evolution to a desired quantum state. Several implementations of QOC, such as GRAPE \cite{GRAPE}, CRAB \cite{CRAB}, and Krotov \cite{Krotov}, have been applied to optimal control calculations of multi-qubit systems. However, the most daunting challenge common to all these QOC approaches is the exponential increase in the Hamiltonian size, which results in concomitant demands in RAM and CPU resources. For instance, recent benchmarks have shown that 128 classical CPUs are required for QOC simulations of 10 qubits, whereas 12 qubits is the current limit for the GRAPE-based algorithm on a quantum-based processor \cite{Lu2017}. Despite the numerous computational techniques used to accelerate QOC calculations \cite{Wang2022,leung2017}, to the best of our knowledge, there have been few efforts to simplify/accelerate QOC simulations of multi-qubit systems that take advantage of their intrinsic symmetry.

In this work, we present a new gradient-based QOC framework for controlling a large system of multiple-entangled qubits. Our approach harnesses the symmetry of finite groups inherent to a large family of Hamiltonians, making it possible to decompose the Hilbert space of the multi-qubit system into computationally tractable subspaces. This new approach allows us to transform the Hamiltonians into block diagonal forms where the evolution of the quantum states is restricted within each symmetry-protected subspace. The transformed Hamiltonians significantly accelerate QOC calculations of multi-qubit dynamics and provide a physically intuitive picture of the selection rules intrinsic to the symmetry of the system. By tuning the bias field and coupling coefficients, transitions in each subspace can be tailored without breaking the intrinsic symmetry of the system. {It is worth mentioning that there has been prior work on decomposing the Hilbert space of permutation-symmetric ($S_n$) multi-qubit systems \cite{bacon2006,albertini2018,albertini2021,dalessandro2023}; however, these previous studies only considered analytical methods of small qubit systems. Our work generalizes this decomposition approach to the dihedral group ($D_n$) symmetry (which brings more controllability to a multi-qubit system than $S_n$ symmetry) and provides a mathematical justification for this approach. Multi-qubit systems with the symmetry of other finite groups can also be analyzed and simplified with our approach. In addition, we provide an open-source Python code (see Data Availability section) and detailed comparisons of execution times between the conventional and symmetry-based methods. We provided numerical analyses for $3$- to $14$-qubit systems, which are much larger than the small quantum systems previously studied with analytical methods.} We also show that our approach can be generalized to nearly all multi-qubit systems by utilizing the Lie-Trotter-Suzuki decomposition scheme \cite{Childs2019, Barthel2020}. Our symmetry-based approach breaks the previous bottleneck of 12 qubits and pushes the limit of QOC calculations to 14 qubits and beyond. {Our work provides proposals for preparing commonly-used symmetric states \cite{chen2017} and realizing simultaneous gate operations \cite{shor1994, grover1996} in symmetry-protected subspaces. Moreover, our work could potentially benefit quantum machine learning studies employing the symmetry of data and quantum circuits \cite{nguyen2022,skolik2023}.}

\section{Results} \label{results}

\subsection{Dynamics and Symmetry of the Multi-Qubit System} \label{dynamics_symmetry}

\begin{figure}[htbp]
\centering
    \begin{subfigure}{0.485\textwidth}
    \centering
    \includegraphics[width=1.0\textwidth]{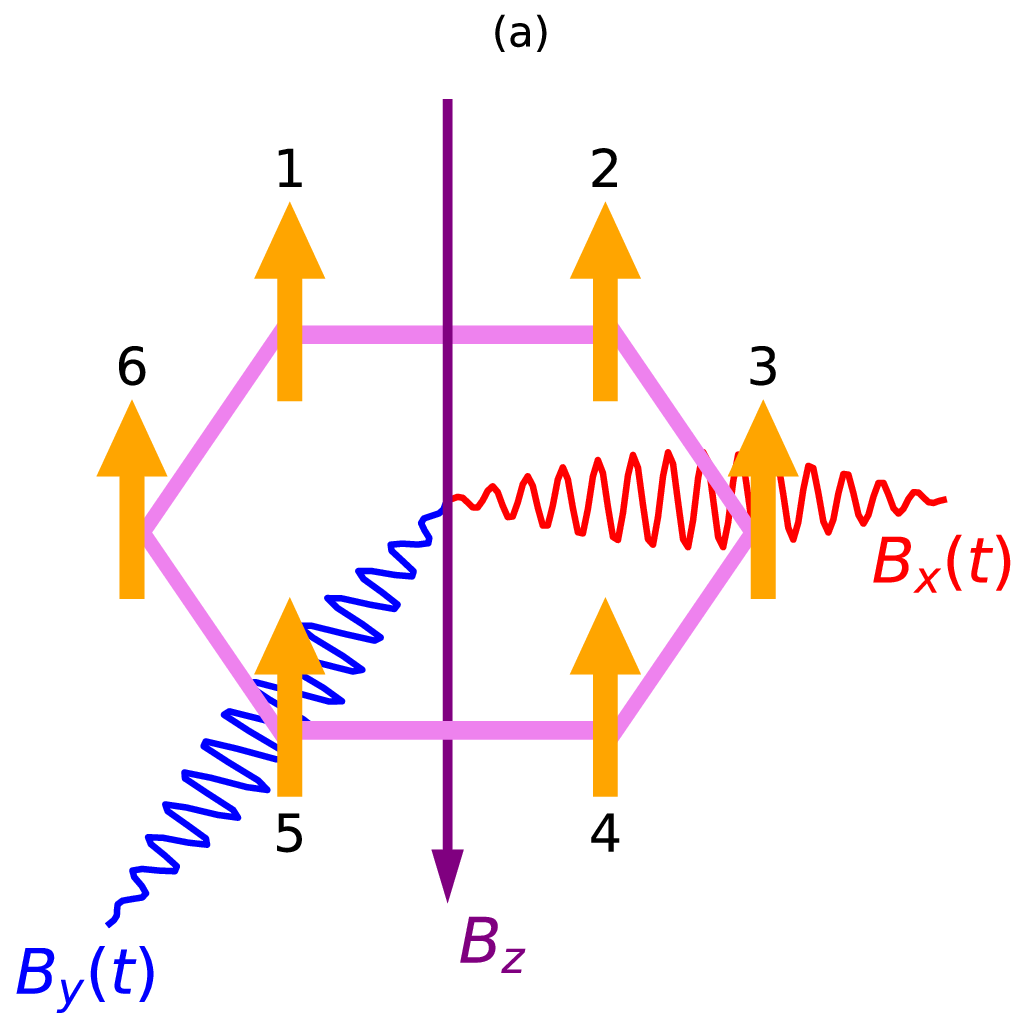}
    \end{subfigure}
    ~
    \begin{subfigure}{0.485\textwidth}
    \centering
    \includegraphics[width=1.0\textwidth]{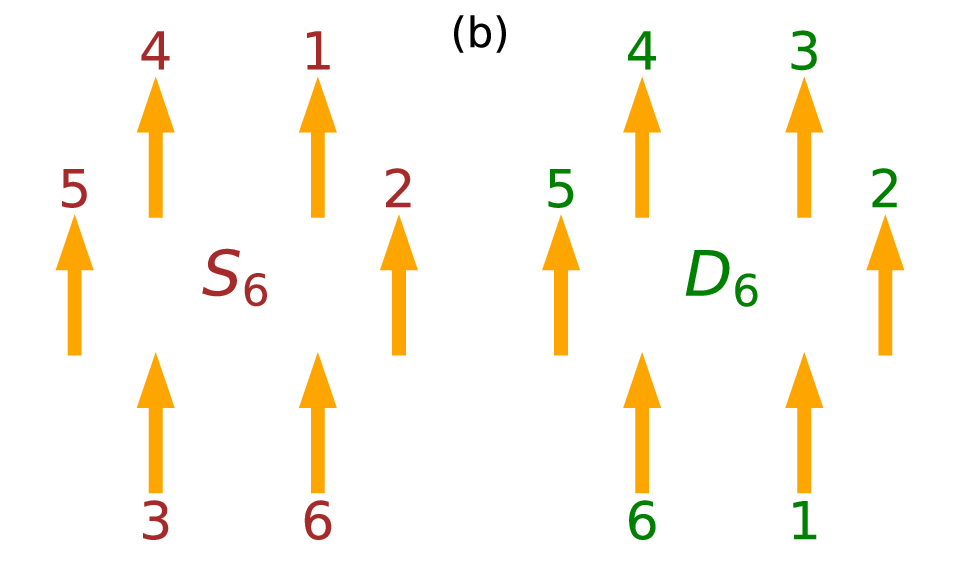}
    \end{subfigure}
    ~
    \begin{subfigure}{0.485\textwidth}
    \centering
    \includegraphics[width=1.0\textwidth]{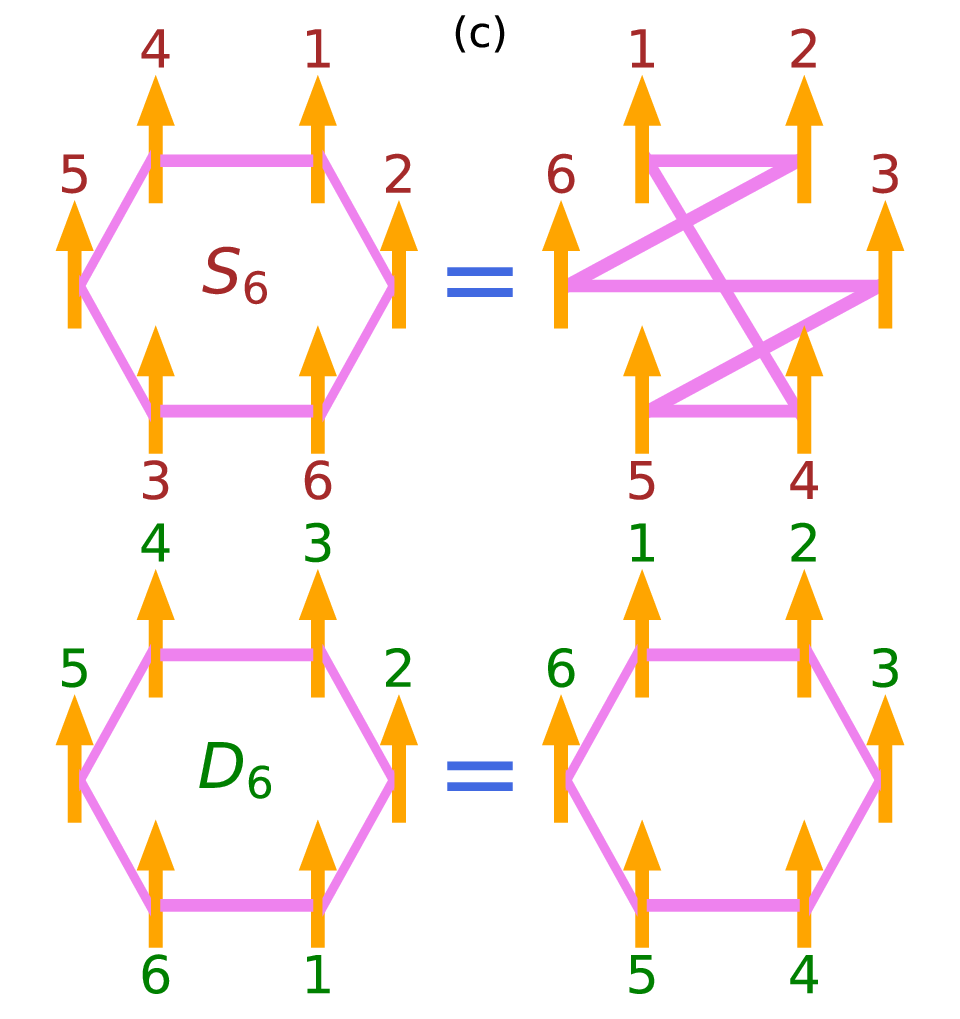}
    \end{subfigure}
\caption{\textbf{Schematic of a multi-qubit system.} (a) A $6$-qubit system in the presence of a static field $B_z$ and time-dependent control pulses $B_x(t)$ and $B_y(t)$. Each qubit is represented by an orange arrow, and the numbers denote the indices of the qubits. The coupling between neighboring qubits is represented by violet bonds. (b) The $6$-qubit system without coupling after applying an $S_6$ action (left) or a $D_6$ action (right) on the indices. (c) The $6$-qubit system with coupling after applying an $S_6$ action (above) or a $D_6$ action (below) on the indices. The configurations connected with an equal sign are equivalent.}
\label{spin_lattice}
\end{figure}

Treating each qubit as a spin-$\frac{1}{2}$ particle, the quantum state $\vert\psi(t)\rangle$ of an $n$-qubit system lies in the $\mathcal{H} (\mathbb{C}^{2^n})$ Hilbert space. The dynamics of a multi-qubit system are governed by the time-dependent Schr\"{o}dinger equation
\begin{equation}
i\frac{\partial}{\partial t}\vert\psi(t)\rangle = \left( H_0+H_c(t) \right) \vert\psi(t)\rangle, \label{schrodingerEq}
\end{equation}
where $H_0$ is the static Hamiltonian and $H_c(t)$ is the time-dependent control Hamiltonian representing the external electromagnetic pulse(s). Fig.~\ref{spin_lattice}a shows a schematic of a $6$-qubit system with nearest-neighbor coupling under a static field along the $z$-axis and time-dependent pulses along the $x$- and $y$-axes. Given an initial state $\vert\psi(0)\rangle$, the final state $\vert\psi(T)\rangle$ can be formally calculated as follows:
\begin{equation}
\vert\psi(T)\rangle = \text{exp}\left(-i \int_0^T \left( H_0+H_c(t) \right) \text{d}t \right)\vert\psi(0)\rangle. \label{solutionschrodingerEq}
\end{equation}
To obtain numerical solutions of Eq.~\ref{solutionschrodingerEq}, we can discretize the control duration $[0, T]$ into $N$ time steps of duration $\tau = \frac{T}{N}$ \cite{GRAPE, Raza2021}. With this approximation, the discrete propagation becomes
\begin{equation}
\vert\psi_{j+1}\rangle = \text{exp} \left( -i \tau \left( H_0+H_c [ (j+\frac{1}{2} )\tau ] \right) \right) \vert\psi_j\rangle,
\label{exppropagator}
\end{equation}
where $\vert\psi_j\rangle$ is the state at time $t = j \tau$. Our QOC framework focuses on optimizing the temporal form of the control pulses to evolve a multi-qubit system to a desired target state $\vert\psi_f\rangle$ (see Sec. IA in the Supplementary Material for details). More precisely, we seek to maximize the probability of the final state $\vert\psi_N\rangle$ being in the desired target state $\vert\psi_f\rangle$ given by
\begin{equation}
P(\vert\psi_N\rangle) = {\vert \langle \psi_f \vert \psi_N \rangle \vert}^2 . \label{probability}
\end{equation}

The symmetry of finite groups of a multi-qubit system arises from the \textit{homogeneity} and \textit{distinguishability} of all the qubits. More specifically, all the qubits can be described by the same Hamiltonians, and each qubit can be distinguished from the others and assigned a unique index. The Hamiltonian of an $n$-qubit system commonly consists of the following terms: $H_z = \sum_{i=1}^{n}\sigma_z^{(i)}$, $H_x = \sum_{i=1}^{n}\sigma_x^{(i)}$, and $H_y = \sum_{i=1}^{n}\sigma_y^{(i)}$ \cite{Lu2017}. We denote $\sigma_\alpha^{(i)}$ as an abbreviation for the $2^n\times2^n$ matrix $\mathbb{I}_2^{\otimes i-1} \otimes \sigma_\alpha \otimes \mathbb{I}_2^{\otimes n-i}$ for $\alpha = x, y, z$. In short, $\sigma_\alpha^{(i)}$ measures the projection of the $i$th qubit's spin along the $\alpha$-axis, where $\sigma_\alpha$ is the Pauli matrix, and $\mathbb{I}_2$ is the rank-$2$ identity matrix. These terms have the symmetry of the permutation group $S_n$ and are, therefore, not affected by any permutation of the qubit indices \cite{ma2007, han1987, li2019}. Note that the group actions are on the indices of the qubits, which does not require repositioning the qubits physically. When the interaction between neighboring qubits in a ring-shaped lattice is considered, one must include the coupling term $H_{z, \text{cpl}} = \sum_{i=1}^{n}\sigma_z^{(i)}\sigma_z^{(i+1)}$ with the boundary condition $\sigma_z^{(n+i)}=\sigma_z^{(i)}$ for $1 \leq i \leq n$. Here $\sigma_z^{(i)}\sigma_z^{(i+1)}$ denotes $\mathbb{I}_2^{\otimes i-1} \otimes \sigma_z \otimes \sigma_z\otimes \mathbb{I}_2^{\otimes n-i-1}$ and represents the coupling between the neighboring $i$th and $i+1$th qubits. The coupling term has the symmetry of the dihedral group $D_n$, and is invariant only under rotations and reflections of the indices of the qubits \cite{ma2007, han1987, li2019}. Note that all terms having $S_n$ symmetry also have $D_n$ symmetry since $D_n$ is a subgroup of $S_n$. Fig.~\ref{spin_lattice}b visually shows that the configuration of the non-interacting qubits is not affected by any $S_n$ or $D_n$ action on the indices. However, when coupling is considered, the system is invariant only under $D_n$ actions (see Fig.~\ref{spin_lattice}c).

\begin{figure}[htbp]
\centering
    \begin{subfigure}{0.305\textwidth}
    \centering
    \includegraphics[width=1.0\textwidth]{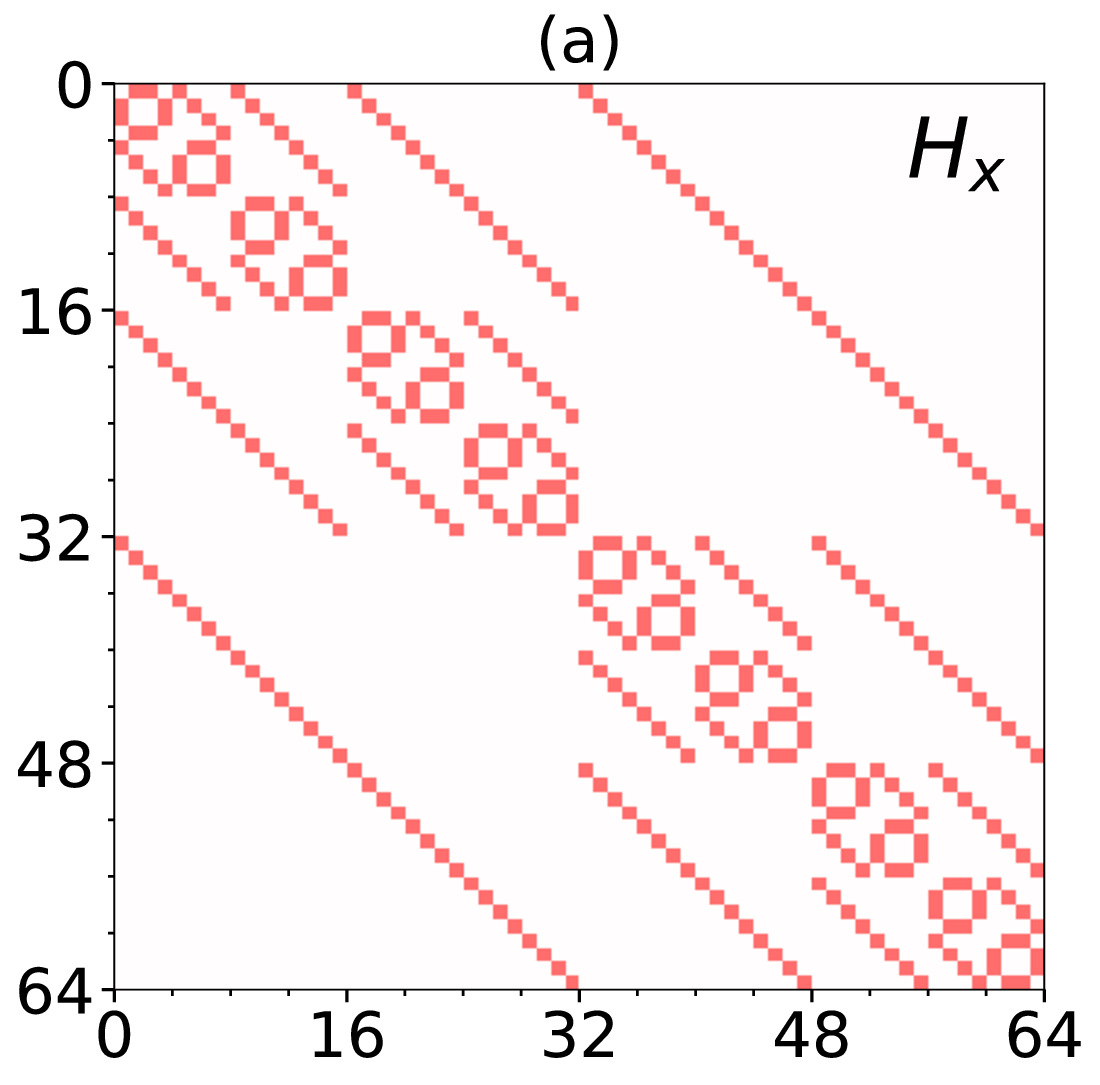}
    \end{subfigure}
    ~
    \begin{subfigure}{0.305\textwidth}
    \centering
    \includegraphics[width=1.0\textwidth]{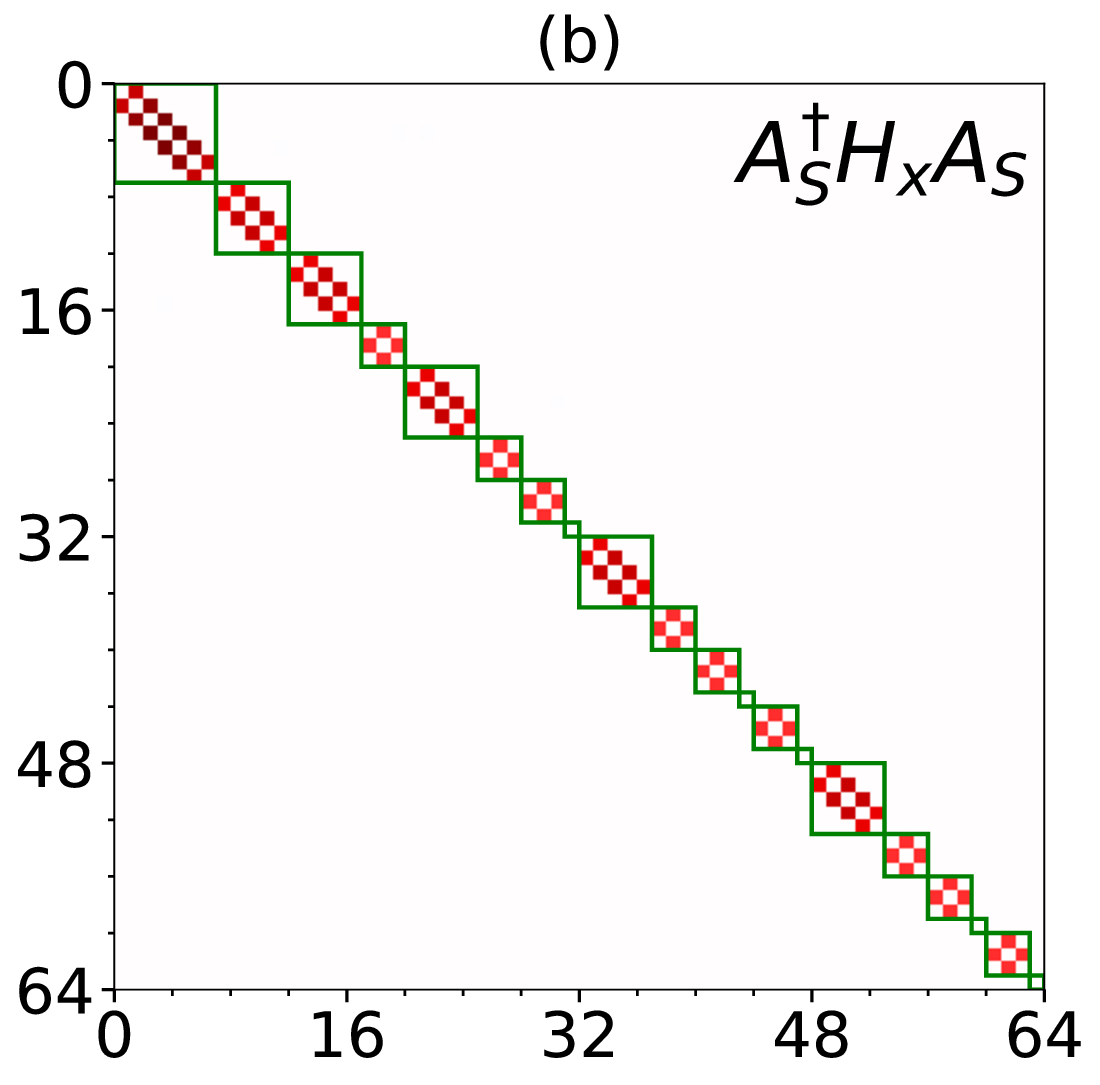}
    \end{subfigure}
    ~
    \begin{subfigure}{0.335\textwidth}
    \centering
    \includegraphics[width=1.0\textwidth]{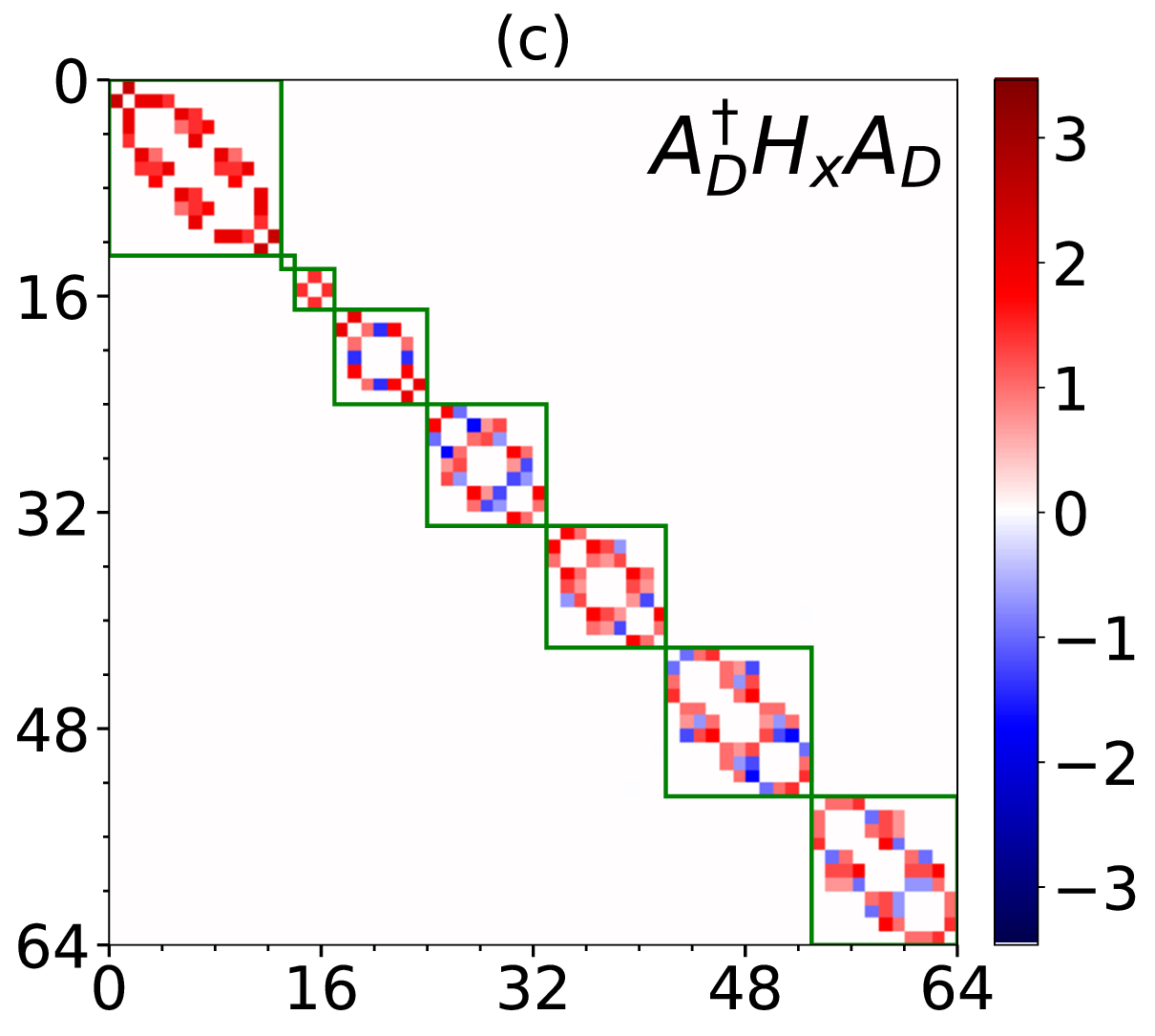}
    \end{subfigure}
    ~
    \begin{subfigure}{0.305\textwidth}
    \centering
    \includegraphics[width=1.0\textwidth]{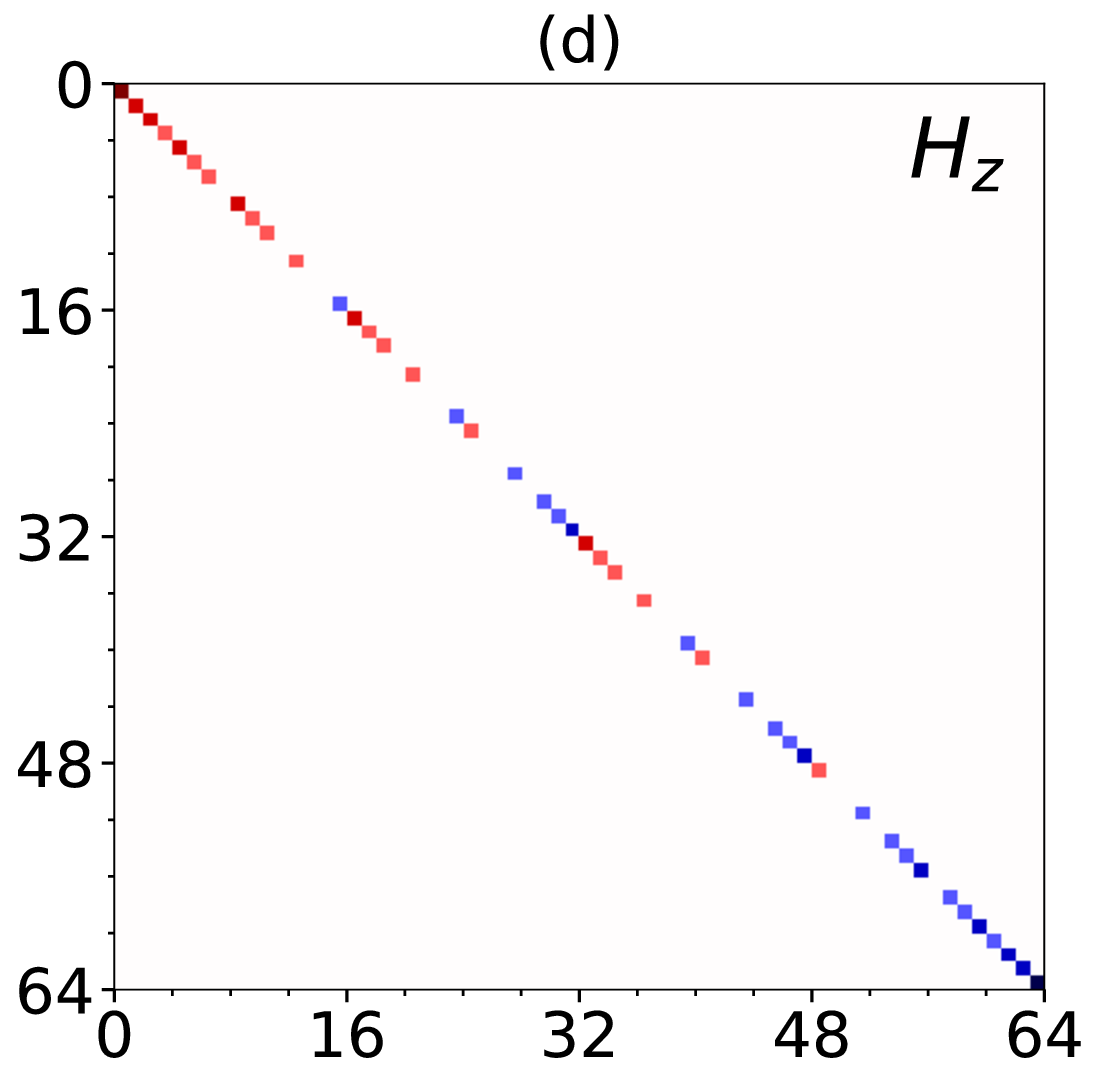}
    \end{subfigure}
    ~
    \begin{subfigure}{0.305\textwidth}
    \centering
    \includegraphics[width=1.0\textwidth]{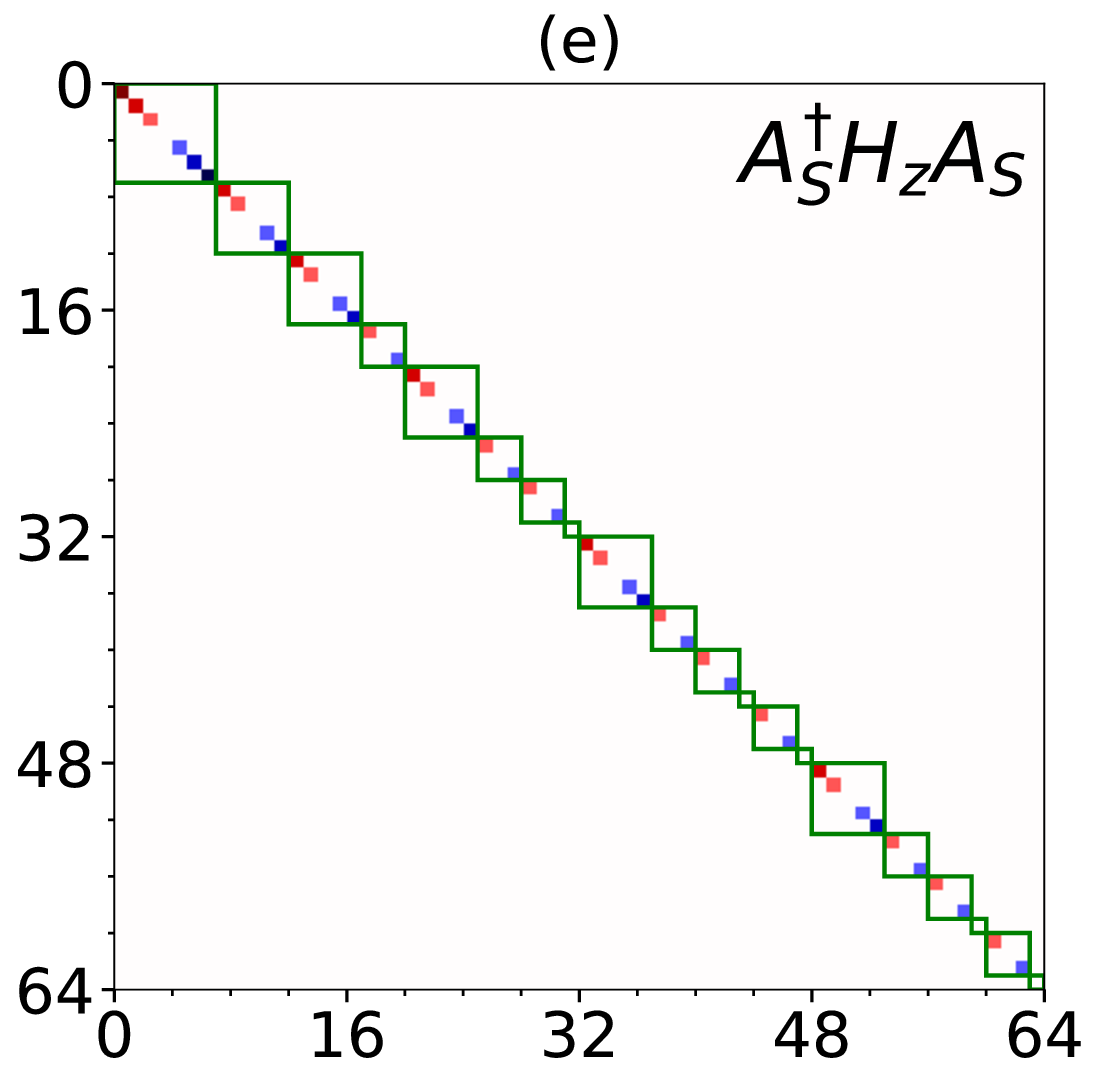}
    \end{subfigure}
    ~
    \begin{subfigure}{0.335\textwidth}
    \centering
    \includegraphics[width=1.0\textwidth]{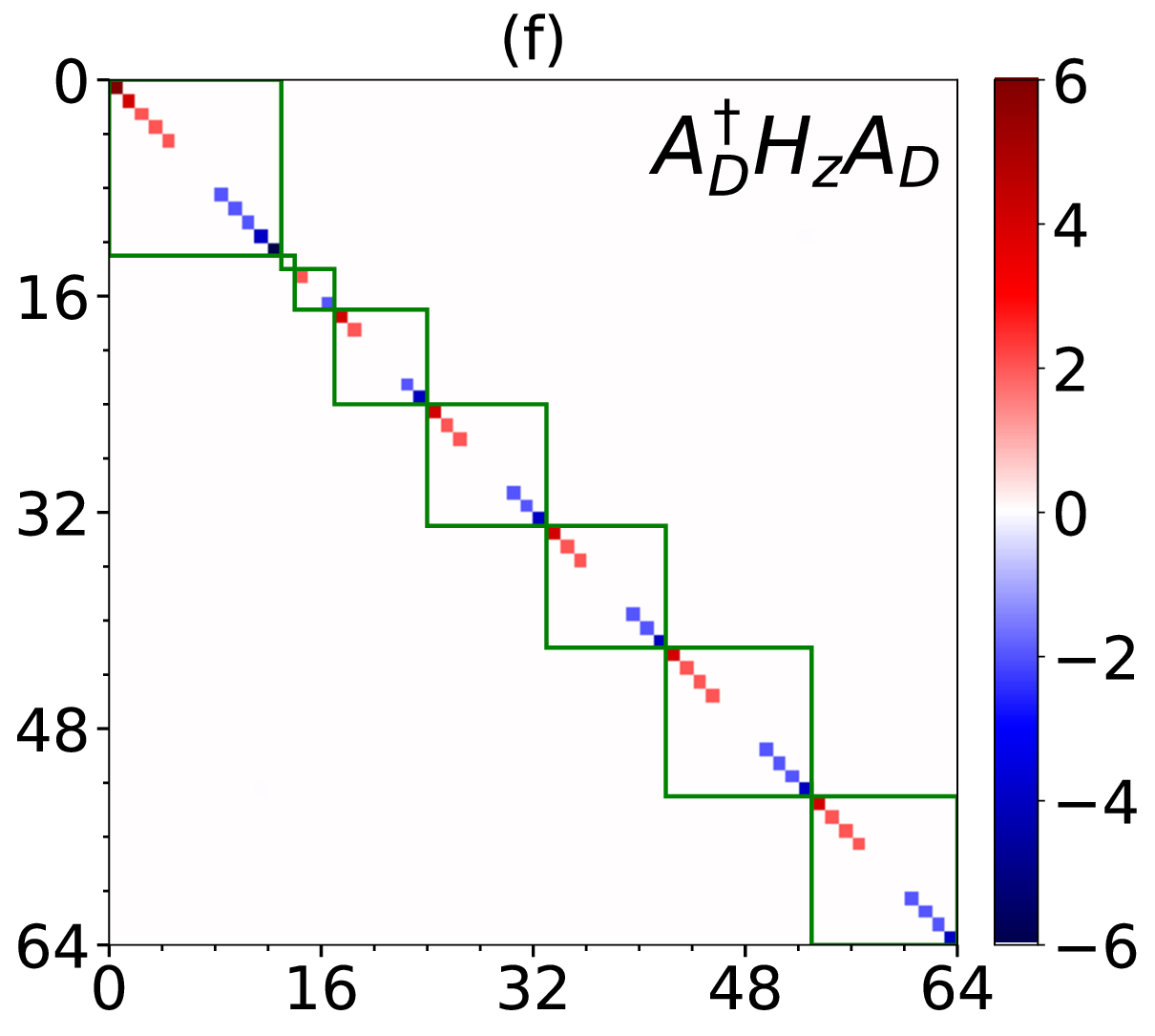}
    \end{subfigure}
    ~
    \begin{subfigure}{0.305\textwidth}
    \centering
    \includegraphics[width=1.0\textwidth]{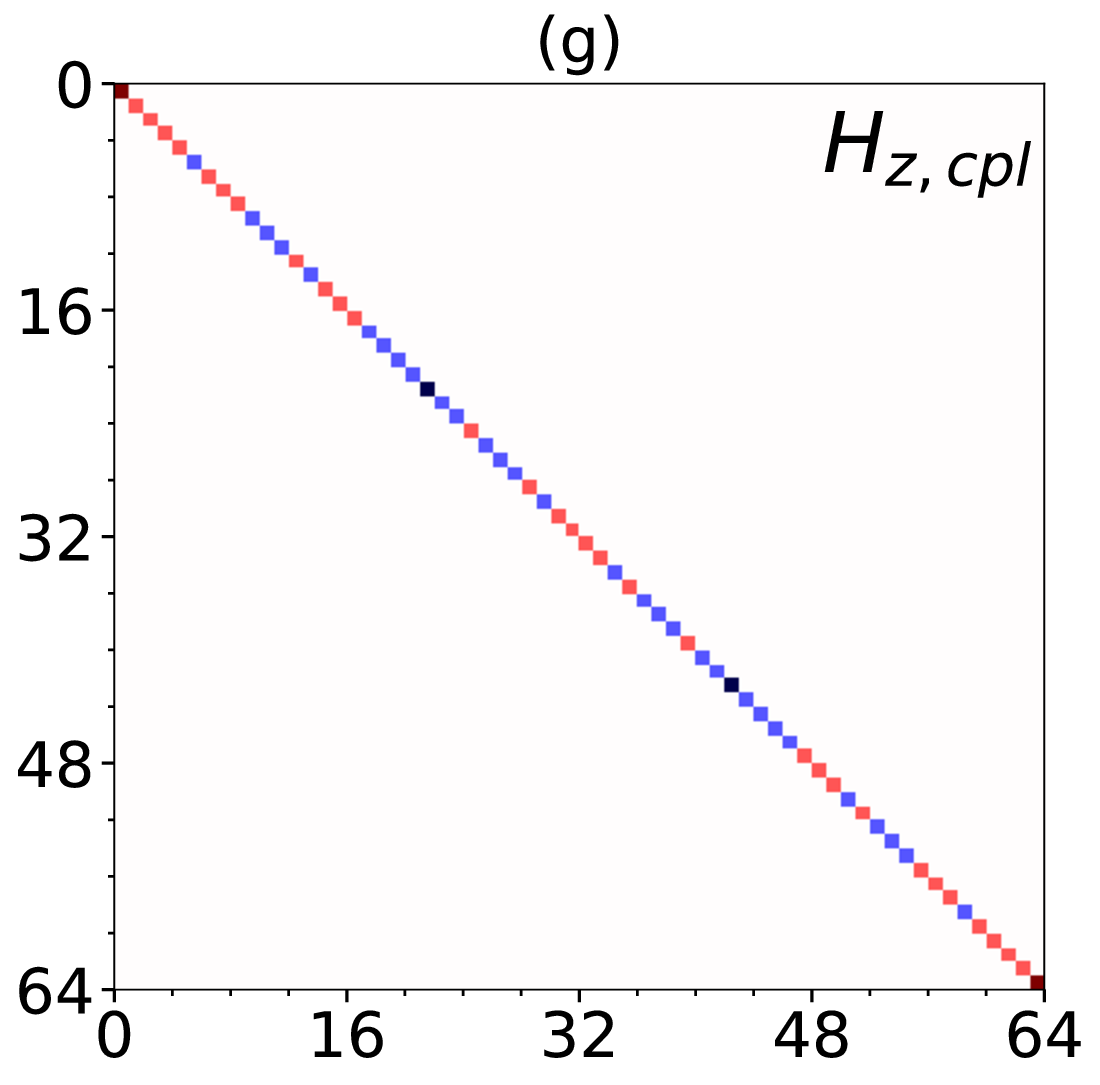}
    \end{subfigure}
    ~
    \begin{subfigure}{0.305\textwidth}
    \centering
    \includegraphics[width=1.0\textwidth]{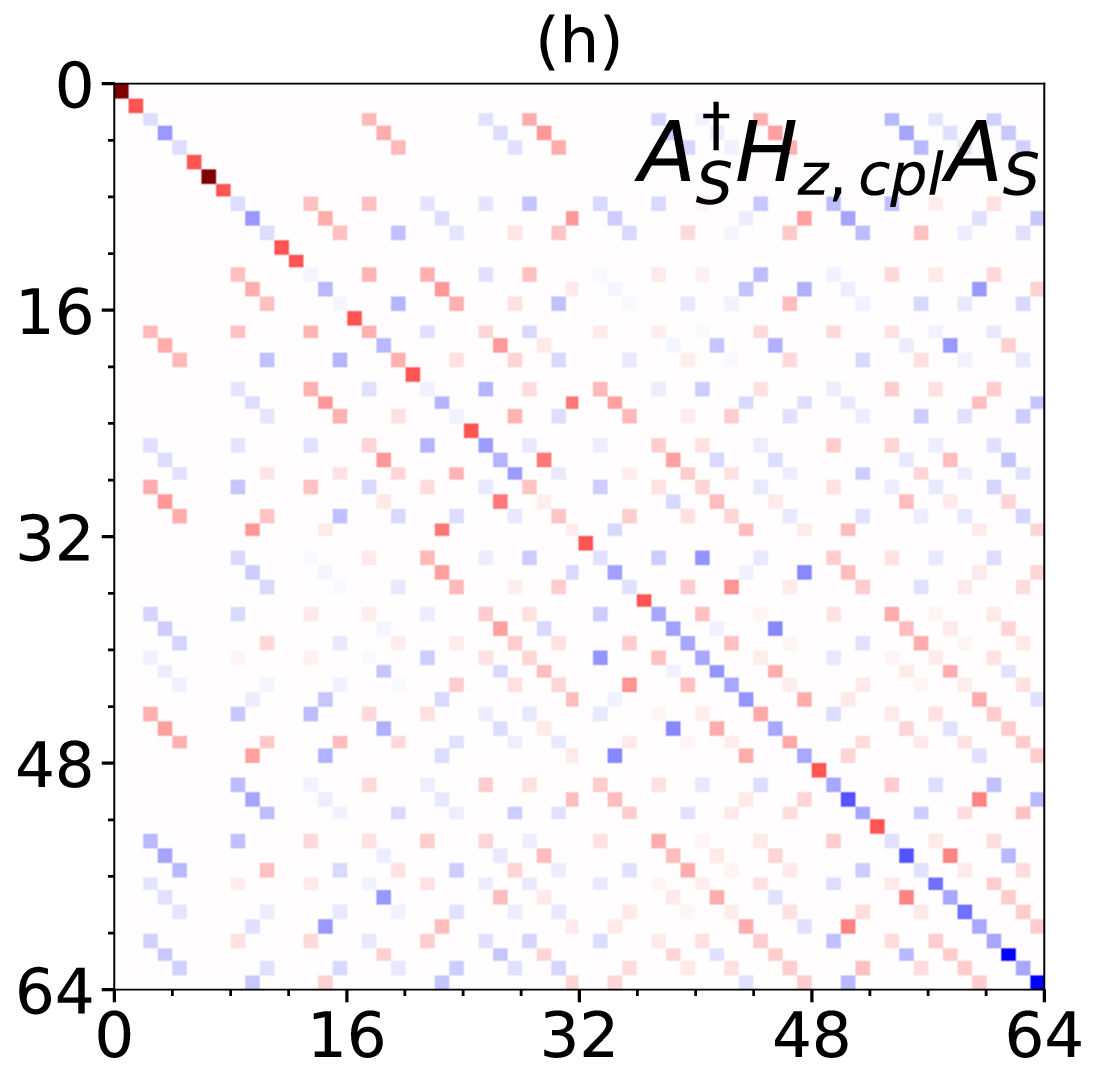}
    \end{subfigure}
    ~
    \begin{subfigure}{0.335\textwidth}
    \centering
    \includegraphics[width=1.0\textwidth]{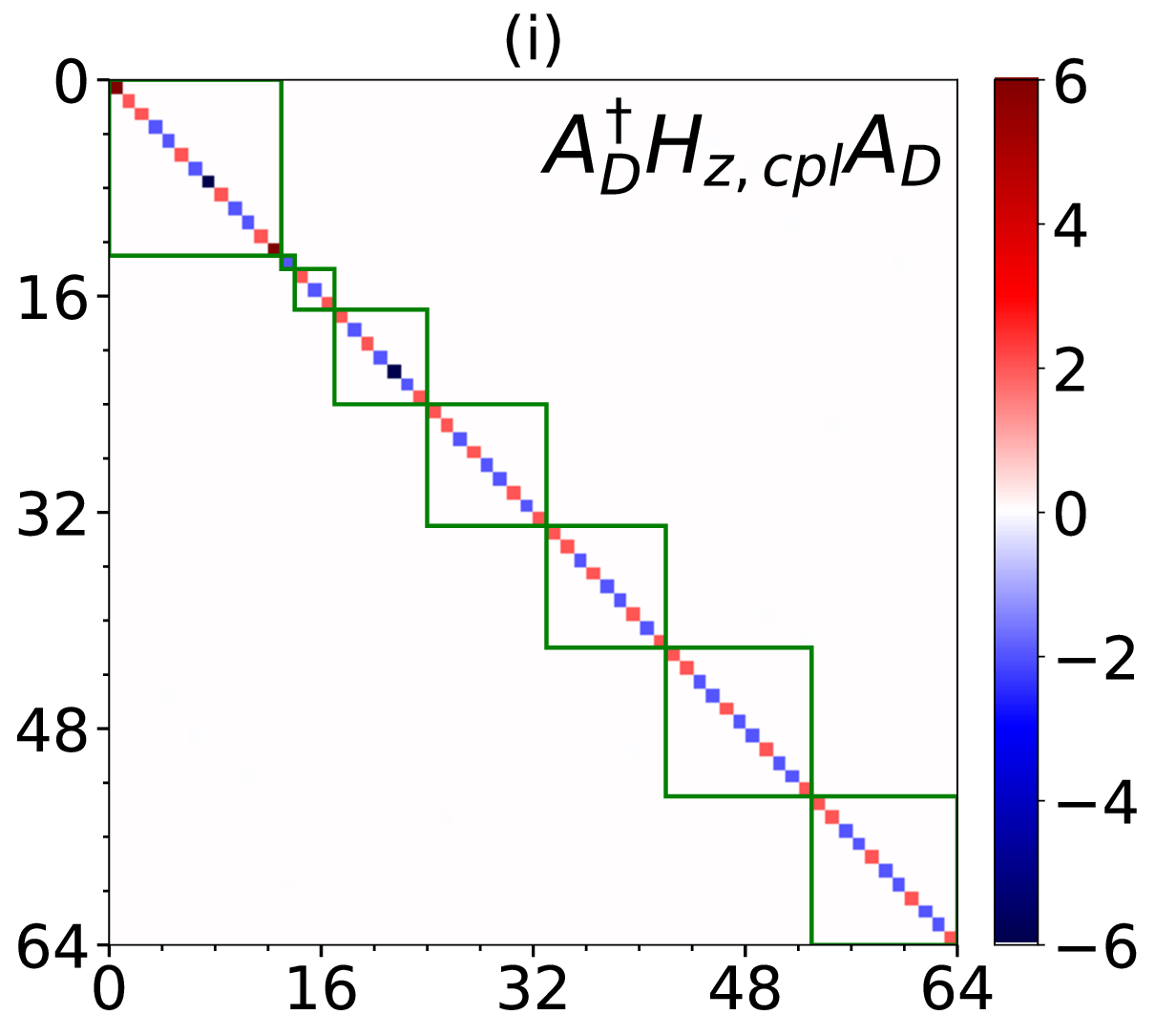}
    \end{subfigure}
\caption{\textbf{Sparsity plots for Hamiltonians of the $6$-qubit system.} (a) $H_x$; (b) $A_S^{\dagger} H_x A_S$; (c) $A_D^{\dagger} H_x A_D$; (d) $H_z$; (e) $A_S^{\dagger} H_z A_S$; (f) $A_D^{\dagger} H_z A_D$; (g) $H_{z,\text{cpl}}$; (h) $A_S^{\dagger} H_{z,\text{cpl}} A_S$; and (i) $A_D^{\dagger} H_{z,\text{cpl}} A_D$. The $x$- and $y$-axes denote the column and row indices of the matrix elements, respectively. The color bars indicate the value of the matrix elements. Each sub-block for the matrices in panels (b), (c), (e), (f), and (i) is enclosed by a green-colored square.}
\label{fig_Hz_Hx_Hzcpl}
\end{figure}

The symmetry of finite groups makes it possible to decompose the Hilbert space $\mathcal{H} (\mathbb{C}^{2^n})$ into orthogonal subspaces. Namely, $\mathcal{H} (\mathbb{C}^{2^n}) = \bigoplus_k \mathcal{H}_k^{S}$ under $S_n$ symmetry or $\mathcal{H} (\mathbb{C}^{2^n}) = \bigoplus_k \mathcal{H}_k^{D}$ under $D_n$ symmetry, where $k \in \mathbb{N}^+$ indexes each specific subspace (see Sec.~\ref{generation_A} and Secs.~IB-E in the Supplementary Material). Under these decompositions, we can find a complete orthogonal basis set in each subspace. The orthogonality and completeness of the subspaces originate from the Schur orthogonality and completeness of the irreducible representations (irreps) of finite groups (see Sec.~IE in the Supplementary Material). \cite{ma2007, han1987, li2019} Putting all the orthonormal bases together as columns, we can construct the adjoint matrix $A$ that transforms the Hamiltonians into \textit{block diagonal} matrices. We denote the $S_n$- and $D_n$-induced adjoint matrices as $A_S$ and $A_D$, respectively. Fig.~\ref{fig_Hz_Hx_Hzcpl} shows that the transformed Hamiltonians $A_S^{\dagger} H_x A_S$ and $A_D^{\dagger} H_x A_D$ are block diagonal, while the original Hamiltonian $H_x$ is not. $A_S^{\dagger} H_z A_S$ and $A_D^{\dagger} H_z A_D$ remain diagonal after transformation, and they follow the same subspace decomposition with $A_S^{\dagger} H_x A_S$ and $A_D^{\dagger} H_x A_D$, respectively. The distribution of the nonzero elements in the complex-valued $H_y$ matrix is the same as those in $H_x$. Lastly, $H_{z, \text{cpl}}$ can be transformed into a block diagonal matrix by only $A_D$ since it does not have $S_n$ symmetry. The sparsity plots of 3-, 4-, 5-, and 7-qubit systems are provided in Figs. S1-S4 in the Supplementary Material. Note that the dimension of a subspace in the Hilbert space decomposition (the subspaces $\mathcal{H}_k^{S}$ or $\mathcal{H}_k^{D}$) is consistent with the size of the corresponding square block in the transformed Hamiltonian (the $k$th green box from top-left to bottom-right in Fig.~\ref{fig_Hz_Hx_Hzcpl}). The first subspace ($\mathcal{H}_1^{S}$ or $\mathcal{H}_1^{D}$) contains the significant ${\vert \uparrow \rangle}^{\otimes n}$ and ${\vert \downarrow \rangle}^{\otimes n}$ states of the multi-qubit system. Table~\ref{dimensionSubspace} shows a comparison between the dimension of the complete Hilbert space and the dimension of the first subspace for systems with various numbers of qubits. While the dimension of $\mathcal{H} (\mathbb{C}^{2^n})$ is $2^n$, the dimensions of $\mathcal{H}_1^{S}$ and $\mathcal{H}_1^{D}$ are reduced to $n+1$ and $\sim O(\frac{2^n}{n})$, respectively.

\begin{table}[htbp]
\begin{center}
\caption{Comparison of the dimensions of $\mathcal{H} (\mathbb{C}^{2^n})$, $\mathcal{H}_1^{S}$, and $\mathcal{H}_1^{D}$} 
\label{dimensionSubspace}
\begin{tabular}{@{}crrr@{}}
\toprule
& \multicolumn{3}{@{}c@{}}{Dimension of space} \\\cmidrule{2-4}
Number of qubits $n$ & $\mathcal{H} (\mathbb{C}^{2^n})$ & $\mathcal{H}_1^{S}$ & $\mathcal{H}_1^{D}$ \\
\midrule
3    & 8  & 4 & 4 \\
4    & 16  & 5 & 6 \\
5    & 32  & 6 & 8 \\
6    & 64  & 7 & 13 \\
7    & 128  & 8 & 18 \\
8    & 256  & 9 & 30 \\
9    & 512  & 10 & 46 \\
10    & 1024  & 11 & 78 \\
11    & 2048  & 12 & 126 \\
12    & 4096  & 13 & 224 \\
13    & 8192  & 14 & 380 \\
14    & 16384  & 15 & 687 \\
\botrule
\end{tabular}
\end{center}
\end{table}

\subsection{Comparison of Conventional and Symmetry-Based Methods} \label{comparison_original_symmetry}

The static Hamiltonian for an Ising model \cite{Lu2017,vstelmachovivc2004,song2019} of a multi-qubit system on a ring-shaped lattice is given by
\begin{equation}
H_0=B_z\cdot\frac{1}{2}\sum_{i=1}^{n}\sigma_z^{(i)}+c_{\text{cpl}}\cdot\frac{1}{4}\sum_{i=1}^{n}\sigma_z^{(i)}\sigma_z^{(i+1)}.
\label{H0WithCoupling}
\end{equation}
The first term represents the interaction between the qubits and a uniform static magnetic field $B_z$ along the $z$-axis. The second term captures the coupling between nearest-neighboring qubits with a strength $c_{\text{cpl}}$. The control Hamiltonian,
\begin{equation}
H_c(t) = B_x(t)\cdot\frac{1}{2}\sum_{i=1}^{n}\sigma_x^{(i)} + B_y(t)\cdot\frac{1}{2}\sum_{i=1}^{n}\sigma_y^{(i)}, \label{Hc}
\end{equation}
manipulates all of the qubits with time-dependent $B_x(t)$ and $B_y(t)$ pulses along the $x$- and $y$-axes. The multi-qubit system has $S_n$ symmetry when $c_{\text{cpl}}$ is zero. When coupling is present, the symmetry of the multi-qubit system reduces to $D_n$. The main goal of our work is to determine the temporal forms of the control pulses, $B_x(t)$ and $B_y(t)$, that excite the multi-qubit system from the initial `all-spin-up' (${\vert \uparrow \rangle}^{\otimes n}$) state to the final `all-spin-down' (${\vert \downarrow \rangle}^{\otimes n}$) state.

\begin{figure*}[htbp]
\centering
    \begin{subfigure}{0.485\textwidth}
    \centering
    \includegraphics[width=1.0\textwidth]{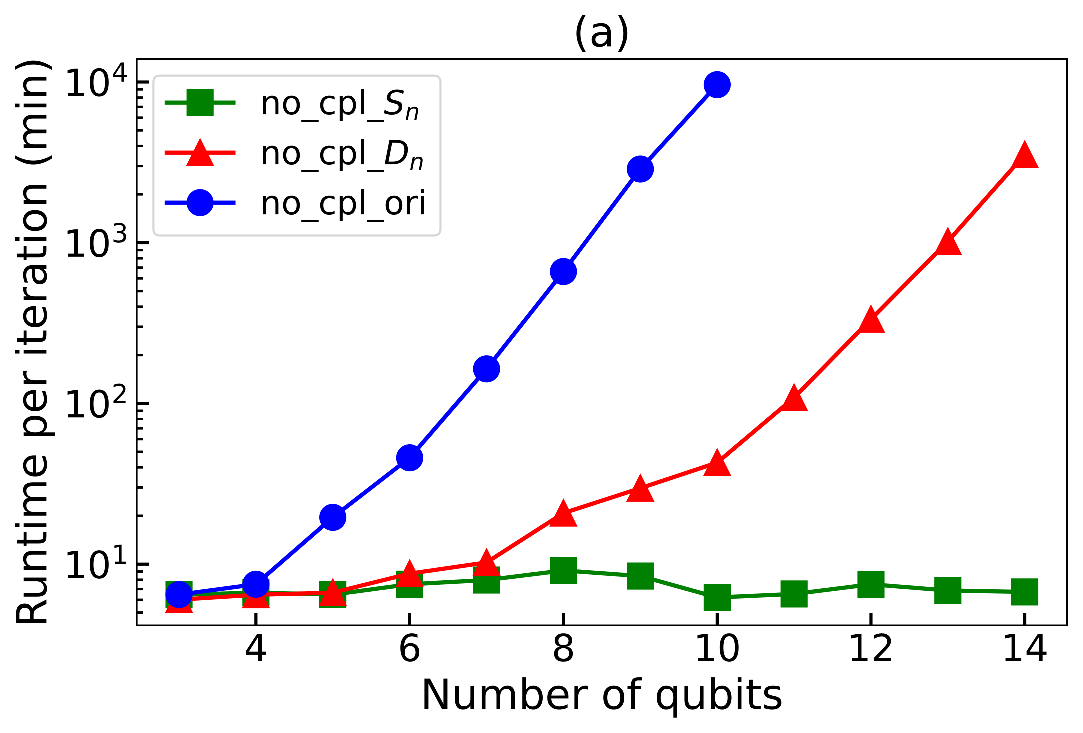}
    \end{subfigure}
    ~
    \begin{subfigure}{0.485\textwidth}
    \centering
    \includegraphics[width=1.0\textwidth]{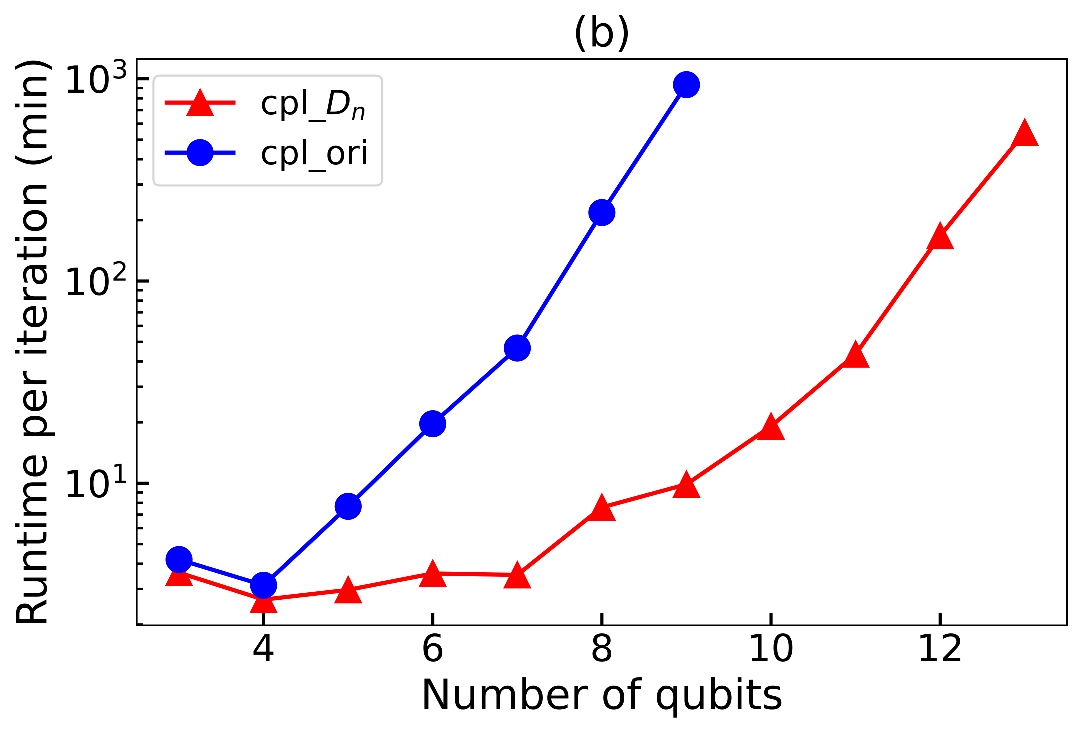}
    \end{subfigure}
    ~
    \begin{subfigure}{0.485\textwidth}
    \centering
    \includegraphics[width=1.0\textwidth]{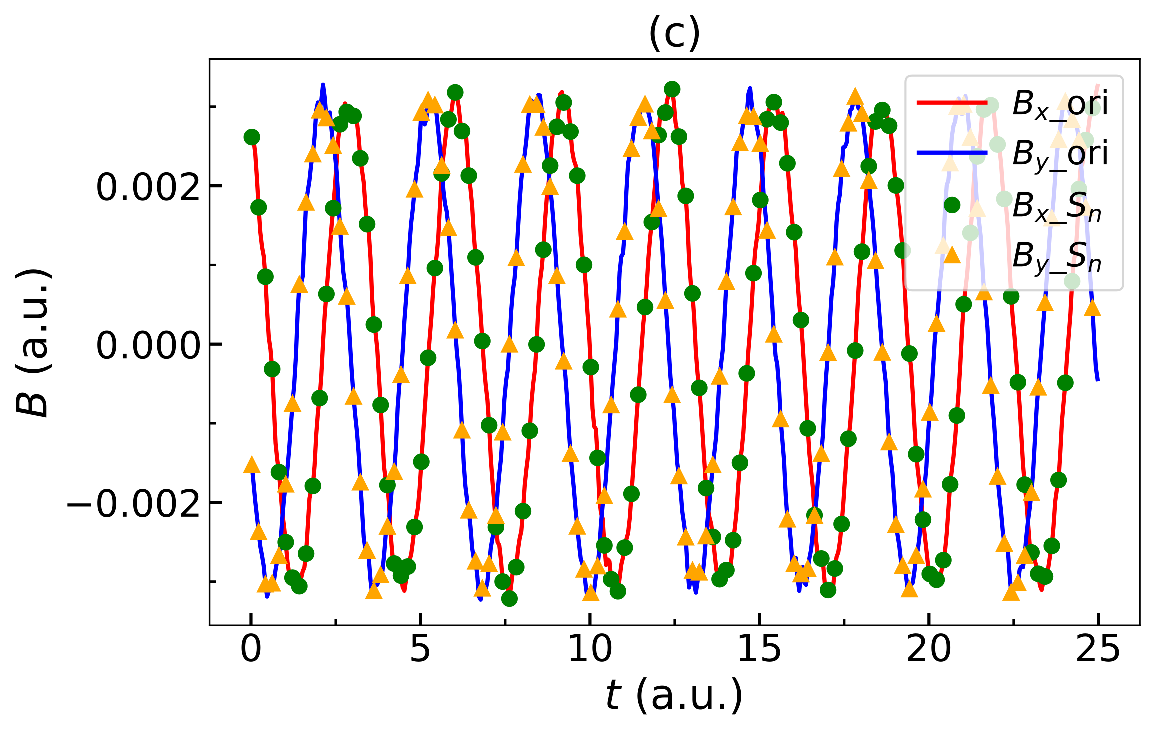}
    \end{subfigure}
    ~
    \begin{subfigure}{0.485\textwidth}
    \centering
    \includegraphics[width=1.0\textwidth]{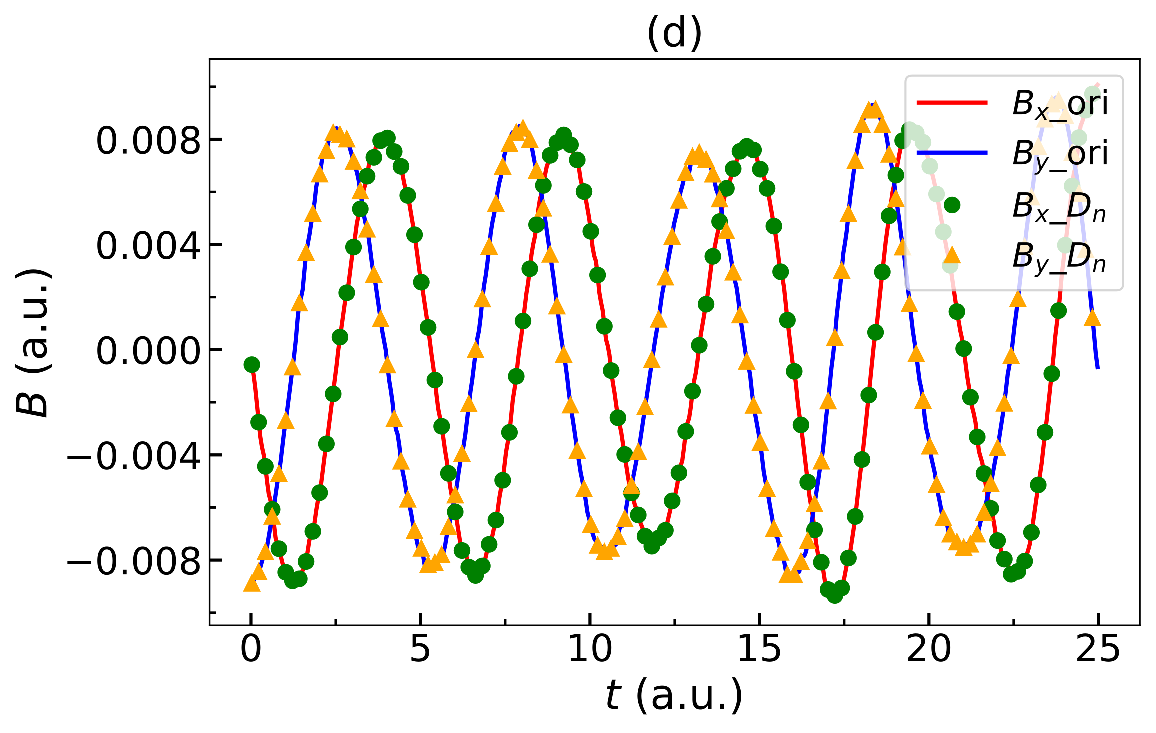}
    \end{subfigure}
    ~
    \begin{subfigure}{0.485\textwidth}
    \centering
    \includegraphics[width=1.0\textwidth]{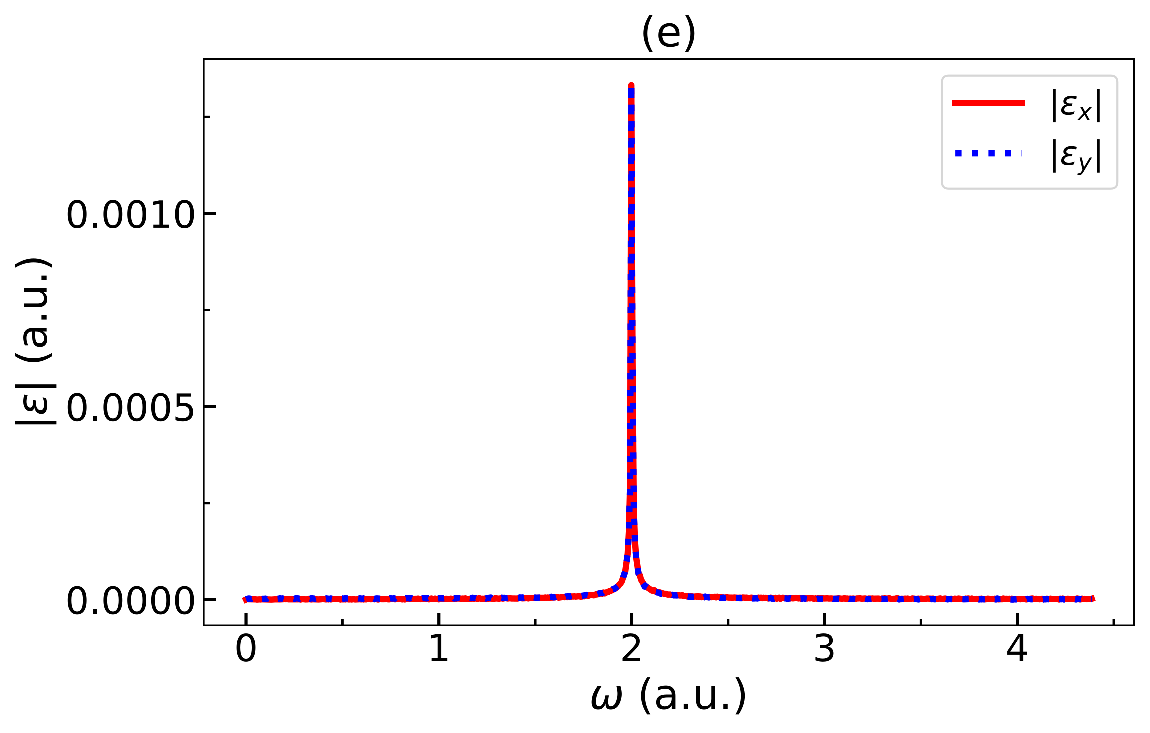}
    \end{subfigure}
    ~
    \begin{subfigure}{0.485\textwidth}
    \centering
    \includegraphics[width=1.0\textwidth]{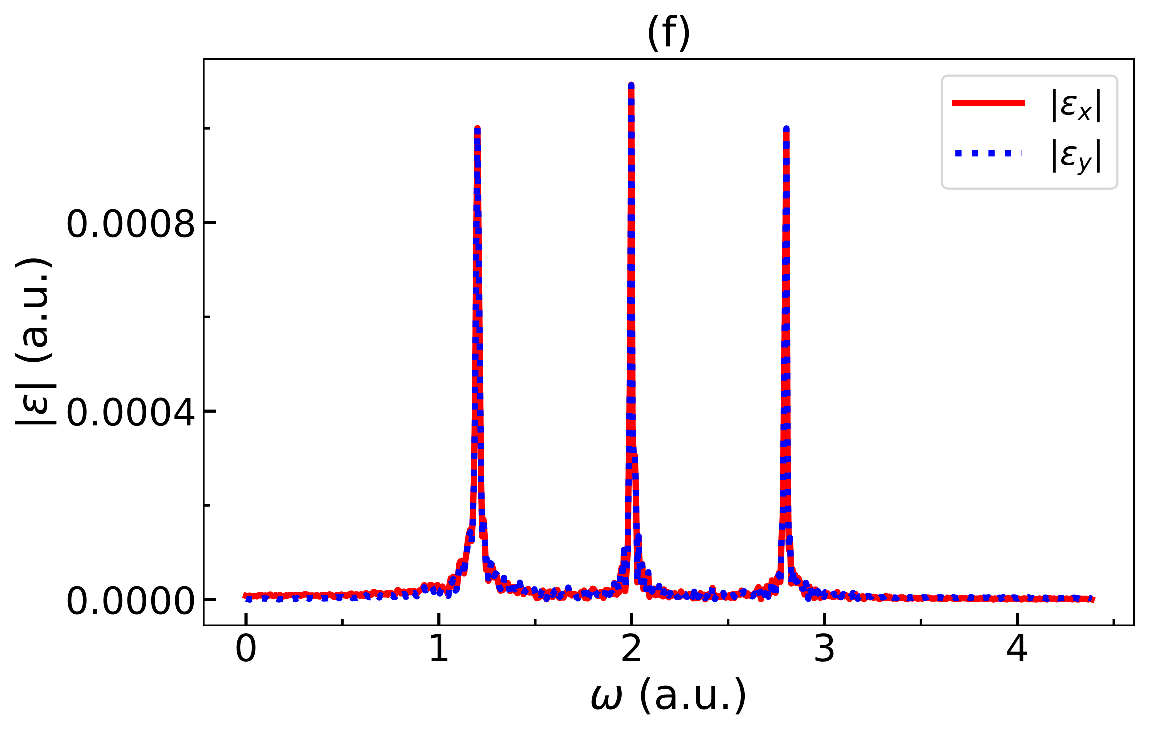}
    \end{subfigure}
\caption{\textbf{Comparison of computational runtimes and optimal control pulses between the conventional and symmetry-based methods.} Runtime comparison between the original and transformed Hamiltonians for systems with (a) no coupling and (b) nearest-neighbor coupling. (c)(d) Optimized pulses $B_x(t)$ and $B_y(t)$, and (e)(f) corresponding power spectra $|\varepsilon(\omega)|$ (i.e., the Fourier transform of the optimized pulses) for a $9$-qubit system with (c)(e) no coupling and (d)(f) nearest-neighbor coupling cases.}
\label{fig_results_nocpl_cpl}
\end{figure*}

When $c_{\text{cpl}}$ is zero, the multi-qubit system is separable, and the evolution of each qubit is independent of any other qubit. In this case, the Hilbert space can be decomposed into the tensor product of $n$ of $2$-dimensional spaces, i.e., $\mathcal{H} (\mathbb{C}^{2^n}) = \bigotimes_{i=1}^n \mathcal{H}^{(i)}(\mathbb{C}^{2})$, and each space $\mathcal{H}^{(i)}(\mathbb{C}^{2})$ can be treated independently. However, in our study, we make use of the direct sum decomposition $\mathcal{H} (\mathbb{C}^{2^n}) = \bigoplus_k \mathcal{H}_k^{S}$ as it is allowed by the symmetry of the $n$-qubit system. As the two states ${\vert \uparrow \rangle}^{\otimes n}$ and ${\vert \downarrow \rangle}^{\otimes n}$ both lie and evolve in the first subspace $\mathcal{H}_1^{S}$, only the first block of $A_S^{\dagger} H_0 A_S$ and $A_S^{\dagger} H_c A_S$ are necessary and sufficient in the calculations. As shown in Fig.~\ref{fig_results_nocpl_cpl}a, the symmetry-based method reduces the runtime by orders of magnitude due to the decreased size of the Hamiltonian from {$2^n \times 2^n$} to {$(n+1) \times (n+1)$}. It should be noted that the separable system with $S_n$ symmetry also has $D_n$ symmetry, and we also compare the computational runtime with the first block of the $A_D$-transformed Hamiltonians. Although it is much more efficient than the original Hamiltonian, the computational runtime is longer than the $A_S$-transformed Hamiltonian since the dimension of $\mathcal{H}_1^{D}$ is larger than $\mathcal{H}_1^{S}$ (see Table~\ref{dimensionSubspace}).

When $c_{\text{cpl}}$ is nonzero, the evolution of each qubit is correlated to the other qubits, and the system is no longer separable, and we cannot use the tensor product decomposition. However, the direct sum decomposition $\mathcal{H} (\mathbb{C}^{2^n}) = \bigoplus_k \mathcal{H}_k^{D}$ can still be leveraged to accelerate the calculation, and we can use the first block of $A_D^{\dagger} H_0 A_D$ and $A_D^{\dagger} H_c A_D$.  Fig.~\ref{fig_results_nocpl_cpl}b shows that compared with the conventional method, the runtime is significantly reduced by the $D_n$-symmetry-based method as well.

Since both $A_S$ and $A_D$ are unitary matrices, the unitary transformation of the Hamiltonians with $A_S$ or $A_D$ does not affect our QOC results. To demonstrate this, we carried out numerical tests for systems ranging from 3 to 14 qubits and found the original and transformed Hamiltonians give exactly the same optimal control pulses $B_x(t)$, $B_y(t)$, and power spectra $|\varepsilon_x(\omega)|$, $|\varepsilon_y(\omega)|$. As an example, Fig.~\ref{fig_results_nocpl_cpl}c and d compare the optimal control pulses for a $9$-qubit system. The data points from our symmetry-based method lie exactly on top of the curves from the conventional method, regardless of whether coupling is present or not. The same comparisons for the 6-, 7-, and 8-qubit systems are given in Fig. S5 in the Supplementary Material. A comparison of the corresponding power spectra is shown in Fig.~\ref{fig_results_nocpl_cpl}e, f. It should be noted that $B_x(t)$ has the same resonance frequency and amplitude as $B_y(t)$ with an additional $\frac{\pi}{2}$ phase shift, which arises from the circular polarization of the control pulses (see Sec.~IF in the Supplementary Material). Fig.~\ref{fig_results_nocpl_cpl}f indicates that the nearest-neighbor coupling terms result in three resonance frequencies in the power spectra. This arises from the energy difference of the transitions in the $\mathcal{H}_1^{S}$ subspace being degenerate when $c_{\text{cpl}} = 0$, whereas the nearest-neighbor coupling terms partially break the degeneracy of the energy differences in the $\mathcal{H}_1^{D}$ subspace.

\section{Discussion} \label{discussion}

\subsection{Symmetry-Protected Subspaces of the Hilbert Space} \label{symmetry_protected_subspace}

We discuss the concept of \textit{symmetry-protected subspaces} inspired by the finite-group-induced decomposition of the Hilbert space $\mathcal{H} (\mathbb{C}^{2^n})$. The system described by Eqs.~\ref{H0WithCoupling} and \ref{Hc} has $S_n/D_n$ symmetry when $c_{\text{cpl}}$ is zero/non-zero. Denoting the state of a single qubit as ${\vert\psi\rangle}_{\text{sq}}$, all the ${\vert\psi\rangle}_{\text{sq}}^{\otimes n}$ states lie in the first subspace $\mathcal{H}_1^{S}/\mathcal{H}_1^{D}$. In fact, $\mathcal{H}_1^{S}$ is a subspace of $\mathcal{H}_1^{D}$. It should be noted that transitions among the ${\vert\psi\rangle}_{\text{sq}}^{\otimes n}$ states can be enabled with quantum gates $U_{\text{sq}}^{\otimes n}$, where $U_{\text{sq}}$ is a single-qubit gate. As such, all $U_{\text{sq}}^{\otimes n}$ transitions are restricted within the specific subspace; i.e., a state in one subspace cannot transition into another subspace as long as the Hamiltonian preserves the symmetry of finite groups. Therefore, we claim that the subspaces generated by decomposing the Hilbert space $\mathcal{H} (\mathbb{C}^{2^n})$ are protected by the symmetry of the finite groups. {Given an initial ${\vert \uparrow \rangle}^{\otimes n}$ state, some important multi-qubit states, such as the Greenberger–Horne–Zeilinger (GHZ) state and the $W$ state \cite{chen2017}, can be realized in the first subspace $\mathcal{H}_1^{S}/\mathcal{H}_1^{D}$. Some essential simultaneous gates in Shor's algorithm for factorizing integers in polylogarithmic time \cite{shor1994} and Grover's algorithm for unstructured search \cite{grover1996}, such as $H^{\otimes n}$ (where $H$ denotes the Hadamard gate), can also be realized in $\mathcal{H}_1^{S}/\mathcal{H}_1^{D}$.}

Physical qubits have been realized in several platforms, such as superconducting qubits \cite{gambetta2017,arute2019}, trapped ions \cite{cirac1995,monz2011}, nitrogen-vacancy centers in diamonds \cite{hensen2015,bradley2019}, and neutral atoms \cite{ebadi2021}. Thus far, all types of physical qubits do not possess an ideal fidelity, which hinders the realization of practical quantum computers. A proposed approach to quantum error correction is to encode one logical qubit with multiple physical qubits. \cite{peres1985,shor1995,fowler2012,google2023} In a symmetry-protected $n$-qubit system, ${\vert\psi\rangle}_{\text{sq}}^{\otimes n}$ states are always in the first subspace regardless of whether the system has $S_n$ symmetry, $D_n$ symmetry, or the symmetry of another subgroup of $S_n$. Therefore, a natural approach is to encode ${\vert \uparrow \rangle}$ and ${\vert \downarrow \rangle}$ with ${\vert \uparrow \rangle}^{\otimes n}$ and ${\vert \downarrow \rangle}^{\otimes n}$, respectively, in the first subspace. We find that the error rate can be greatly reduced not only because the logical qubit is $n$-fold encoded but also because the first subspace is protected by the symmetry of the finite group. In short, the quantum error is significantly suppressed since the quantum state cannot evolve to other subspaces even if the control pulses deviate from the optimized amplitude, resonance frequency, or duration.

\begin{figure*}[htbp]
\centering
    \begin{subfigure}{0.315\textwidth}
    \centering
    \includegraphics[width=1.0\textwidth]{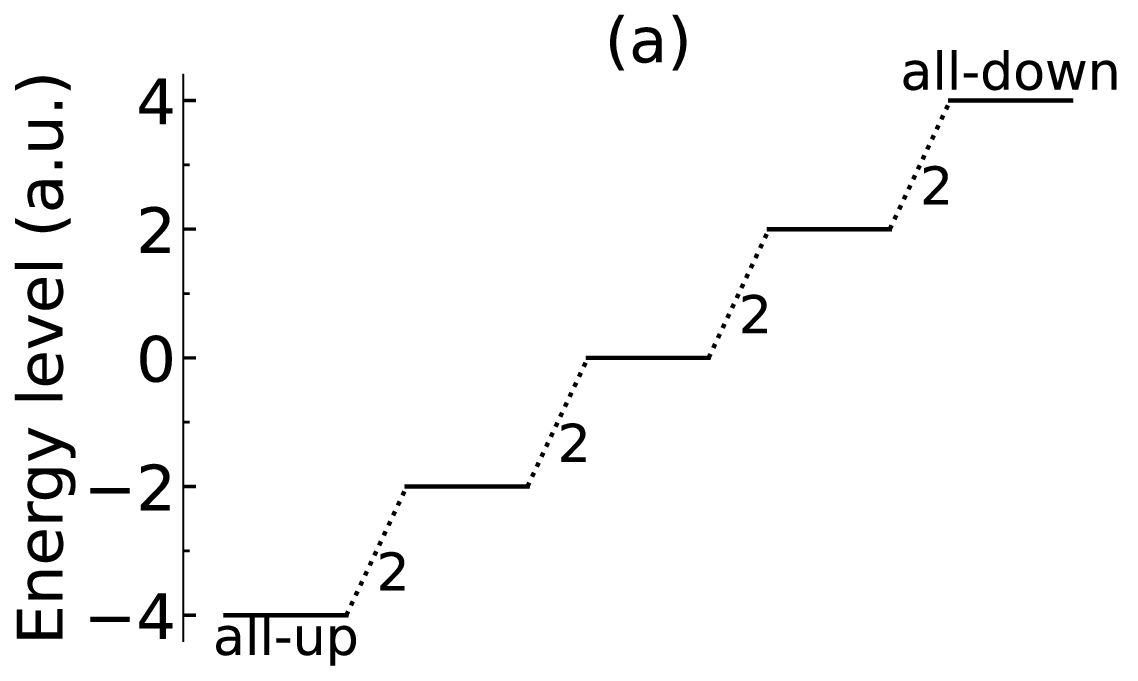}
    \end{subfigure}
    ~
    \begin{subfigure}{0.315\textwidth}
    \centering
    \includegraphics[width=1.0\textwidth]{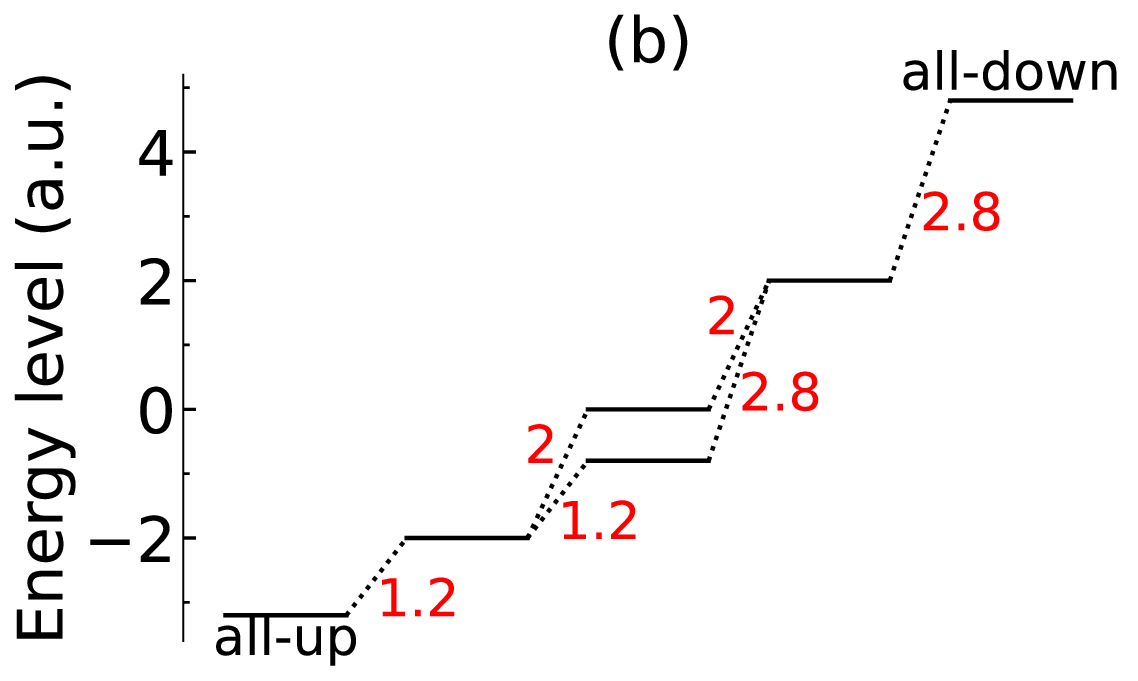}
    \end{subfigure}
    ~
    \begin{subfigure}{0.315\textwidth}
    \centering
    \includegraphics[width=1.0\textwidth]{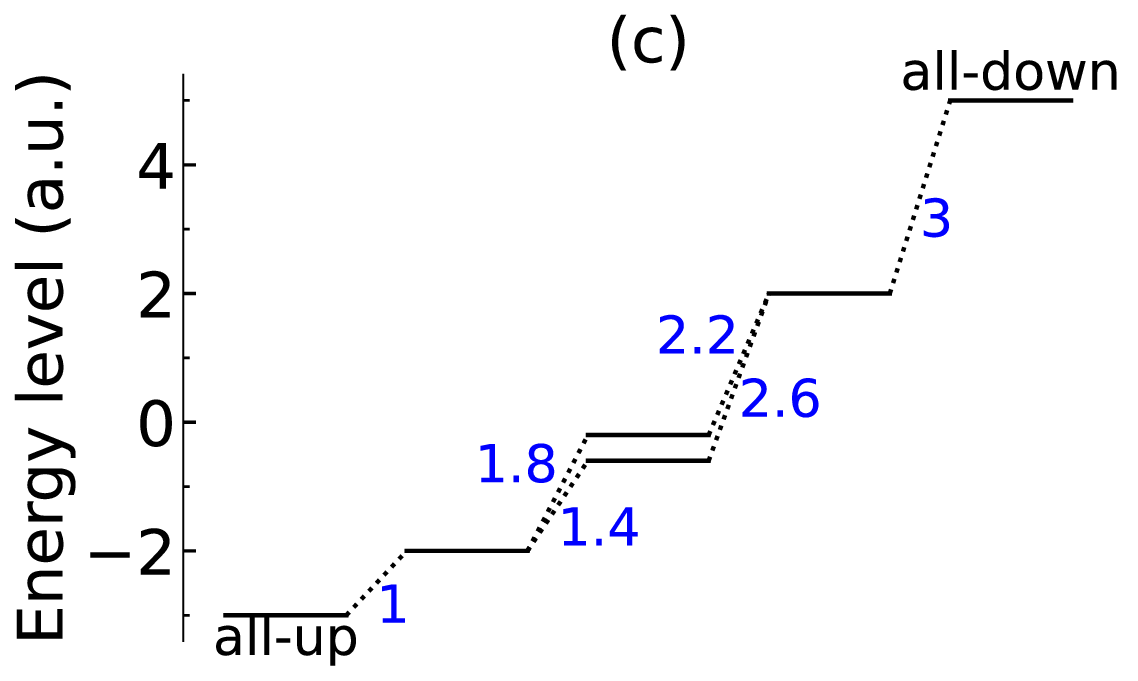}
    \end{subfigure}
    ~
    \begin{subfigure}{0.485\textwidth}
    \centering
    \includegraphics[width=1.0\textwidth]{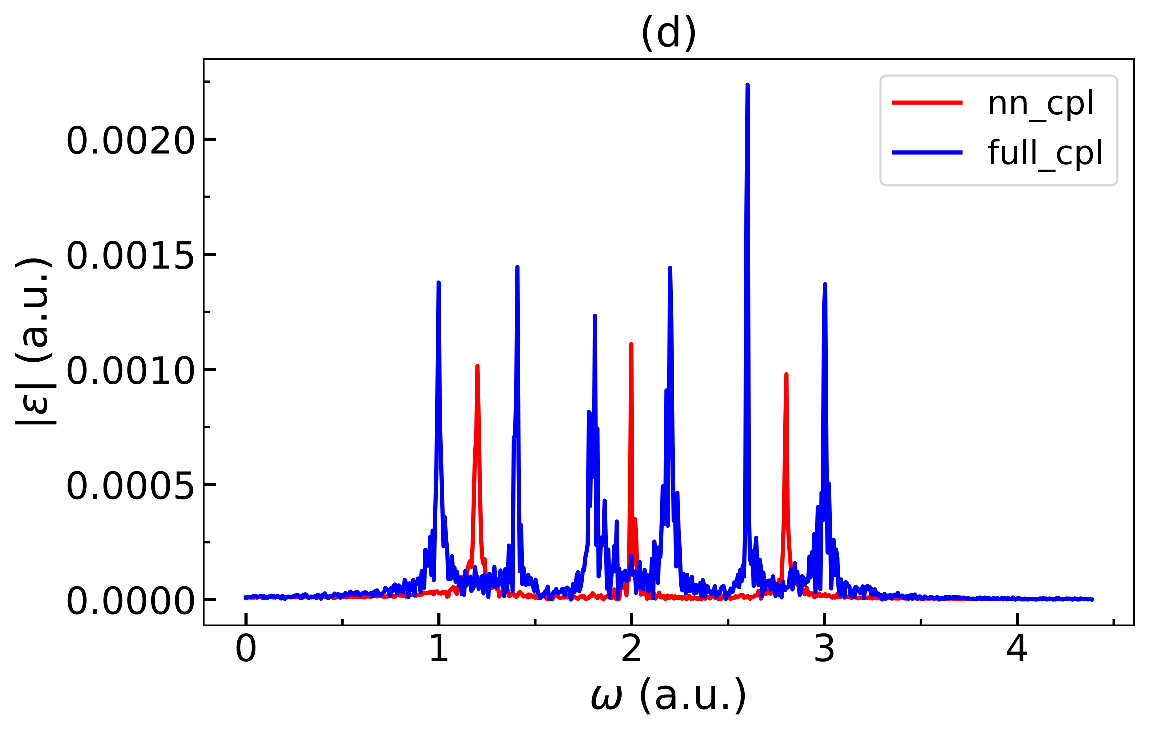}
    \end{subfigure}
    ~
    \begin{subfigure}{0.485\textwidth}
    \centering
    \includegraphics[width=1.0\textwidth]{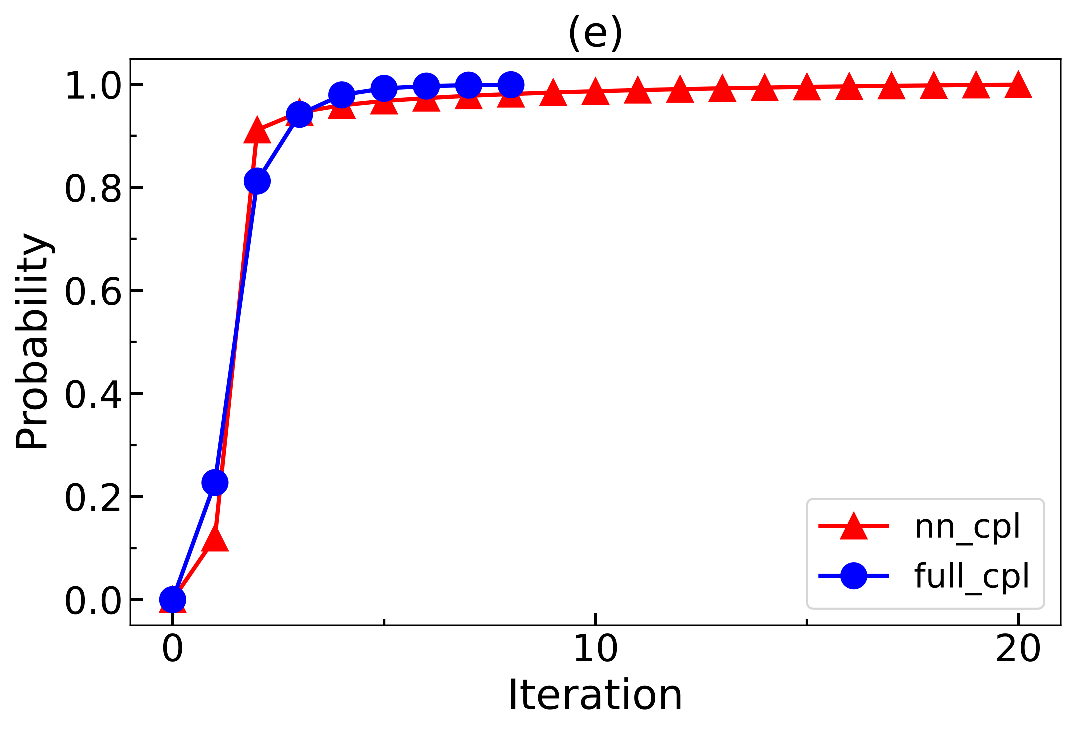}
    \end{subfigure}
\caption{\textbf{Comparison of energy levels, power spectra, and convergence for the systems with nearest-neighbor coupling and full coupling.} Eigenstates in the first subspace of a $4$-qubit system when the system has (a) $S_n$ symmetry without coupling terms, (b) $D_n$ symmetry with nearest-neighbor coupling terms as described in Eq.~\ref{H0WithCoupling}, and (c) $D_n$ symmetry with full coupling as described in Eq.~\ref{H0MoreCoupling1}. The `all-up' and the `all-down' eigenstates at the two ends of the transition cascade are labeled. The transitions permitted by the selection rules are indicated by dashed lines, and the energy differences are shown next to each transition. (d) Power spectra of the optimized pulses when the $4$-qubit system has nearest-neighbor and full coupling. (e) Comparison of convergence for the system with nearest-neighbor and full coupling.}
\label{fig_energy_level_coupling_comparison}
\end{figure*}

Turning our attention to the first subspace, when $c_{\text{cpl}}$ is zero, $n+1$ eigenstates exist in $\mathcal{H}_1^{S}$ with equally-spaced energy levels, as shown in Fig.~\ref{fig_energy_level_coupling_comparison}a. Thus, there is only one resonance frequency, which corresponds to the single peak in Fig.~\ref{fig_results_nocpl_cpl}e. It should be noted that a direct transition is not possible from $\vert\uparrow\rangle^{\otimes n}$ to $\vert\downarrow\rangle^{\otimes n}$ due to selection rules (see SI Sec.~IF). Such a transition can only be realized via a cascade consisting of multiple intermediate eigenstates. Since there is only one resonance frequency in the transition cascade, any pulse exciting one transition in the cascade also enables all other transitions. As a result, given the initial state is $\vert\uparrow\rangle^{\otimes n}$, the only possible final eigenstate is $\vert\downarrow\rangle^{\otimes n}$, which is realized by the gate $\sigma_x^{\otimes n}$, and vice versa. It is, therefore, not possible to evolve the system to any intermediate eigenstate because such a transition cannot be realized by any gate in the form of $U_{\text{sq}}^{\otimes n}$.

To fully control transitions in the first subspace, we can introduce coupling terms to break the degeneracy of the resonance frequencies. As shown in Fig.~\ref{fig_results_nocpl_cpl}f, using the nearest-neighbor coupling in Eq.~\ref{H0WithCoupling}, there will be three resonance frequencies in the $n$-qubit system when $n \geq 3$. However, when $n \geq 4$, three resonance frequencies are insufficient to completely break the degeneracy of the energy differences, as shown in Figs.~\ref{fig_energy_level_coupling_comparison}b and d. Beyond the nearest neighbors, we can introduce further couplings between qubit pairs:
\begin{align}
\begin{aligned}
H_0 &= B_z\cdot\frac{1}{2}\sum_{i=1}^{n}\sigma_z^{(i)} + c_{\text{cpl}}^{(1)}\cdot\frac{1}{4}\sum_{i=1}^{n}\sigma_z^{(i)}\sigma_z^{(i+1)}+ c_{\text{cpl}}^{(2)}\cdot\frac{1}{4}\sum_{i=1}^{n}\sigma_z^{(i)}\sigma_z^{(i+2)} \\
&+ \dots + c_{\text{cpl}}^{(\lfloor \frac{n}{2} \rfloor)}\cdot\frac{1}{4}\sum_{i=1}^{n}\sigma_z^{(i)}\sigma_z^{(i+\lfloor \frac{n}{2} \rfloor)},
\label{H0MoreCoupling1}
\end{aligned}
\end{align}
where $c_{\text{cpl}}^{(1)}$, $c_{\text{cpl}}^{(2)}$, and $c_{\text{cpl}}^{(\lfloor \frac{n}{2} \rfloor)}$ are the nearest-, next-nearest-, and furthest-neighbor coupling strengths, respectively. This form can fully break the degeneracy of energy levels and energy differences. $D_n$ symmetry is preserved with the full coupling terms, and therefore, the eigenstates do not change, whereas their energy levels are modified. As shown in Fig.~\ref{fig_energy_level_coupling_comparison}c, the degeneracy in the energy differences is completely broken in the $4$-qubit system, resulting in $6$ resonance frequencies in the power spectra of the fully coupled system in Fig.~\ref{fig_energy_level_coupling_comparison}d. As such, the cascade of transitions from $\vert\uparrow\rangle^{\otimes n}$ to $\vert\downarrow\rangle^{\otimes n}$ becomes a series of concatenated two-level systems, and each two-level transition can be enabled by pulses of a unique resonance frequency. \cite{cheng2006} This allows us to manipulate the system to be in any eigenstate, or a linear combination of the eigenstates, with a selected route of transitions from the $\vert\uparrow\rangle^{\otimes n}$ initial state (as long as high-quality pulses with desired resonance frequencies and profiles can be generated). In summary, properly tuning the coupling coefficients $c_{\text{cpl}}^{(i)}, 1 \leq i \leq \lfloor \frac{n}{2} \rfloor$ in an $n$-qubit system can completely break the degeneracy of energy differences. The role of each resonance frequency is apparent in a completely non-degenerate system since each one corresponds to an exact transition in the excitation cascade pathway. This enables a more efficient way to manipulate a multi-qubit system. As a demonstration,  Fig.~\ref{fig_energy_level_coupling_comparison}e shows that the probability $P$ in Eq.~\ref{probability} converges in fewer iterations when the $4$-qubit system is fully coupled. The same comparison for 5- and 6-qubit systems is given in Fig. S6 in the Supplementary Material.

We propose that a subspace of the fully coupled multi-qubit system can potentially be a platform for simulating the Hamiltonians of other quantum systems. \cite{manin1980,feynman2018} In a coupled $n$-qubit system, there are $O(\frac{2^n}{n})$ eigenstates in the first subspace under $D_n$ symmetry. With full coupling, their energy levels can be manipulated by tuning $\lfloor \frac{n}{2}\rfloor+1$ parameters, namely the static field $B_z$ and the coupling coefficients $c_{\text{cpl}}^{(i)}$. The transitions in the first subspace can be controlled by pulses with selected resonance frequencies, which enables us to examine the features of the Hamiltonian through the evolution of the multi-qubit system. \cite{lloyd1996,berry2007} Moreover, we can tailor the ``route'' of transitions in the cascade when the degeneracy of resonance frequencies is broken; i.e., even though the selection rules indicate allowed transitions, some undesired transitions can be avoided by filtering the corresponding resonance frequency component in the pulses. This allows more controllability in simulating the Hamiltonian with a subspace of the multi-qubit system.

\subsection{Generalizing the Symmetry-Based Method with the Lie-Trotter-Suzuki Decomposition} \label{LTS_decomposition}

We explore the generalization of our symmetry-based transformation method to other multi-qubit systems with the Lie-Trotter-Suzuki decomposition, or Trotterization, of the propagators. Trotterization is a decomposition that approximates the exponential of a summed-up operator with the product of the exponential of each element in the sum. \cite{Childs2019,Barthel2020} Consider the following control Hamiltonian,
\begin{equation}
H'_c(t) = \frac{1}{2}\sum_{i=1}^{n} \left( B_x^{(i)}(t) \cdot \sigma_x^{(i)} + B_y^{(i)}(t) \cdot \sigma_y^{(i)} \right),
\label{Hcprime}
\end{equation}
where each qubit is tuned by a different control pulse. This Hamiltonian cannot be block diagonalized with the $A_S$-transformation because the $S_n$ symmetry of the $n$-qubit system is broken. However, each term in the sum, i.e., $H_c^{\prime (i)}(t)=B_x^{(i)}(t) \cdot \sigma_x^{(i)} + B_y^{(i)}(t) \cdot \sigma_y^{(i)}$ satisfies $S_1$ symmetry since the control pulse interacts with the $i$th qubit only. {When the static Hamiltonian has no coupling terms, we can calculate the evolution of each qubit separately in the Hilbert space $\mathcal{H} (\mathbb{C}^{2})$. As such, the Hilbert space can be decomposed from $\mathcal{H} (\mathbb{C}^{2^n})$ to ${\mathcal{H} (\mathbb{C}^{2})}^{\otimes n}$, which simplifies the QOC calculation.} For the case of the inseparable system with coupling, the complete Hamiltonian becomes
\begin{equation}
H_0 + H'_c(t) = \frac{1}{2}\sum_{i=1}^{n} \left( B_z \cdot \sigma_z^{(i)} + B_x^{(i)}(t) \cdot \sigma_x^{(i)} + B_y^{(i)}(t) \cdot \sigma_y^{(i)} \right) + c_{\text{cpl}}\cdot\frac{1}{4} \sum_{i=1}^{n}\sigma_z^{(i)}\sigma_z^{(i+1)},
\label{completeH_cpl}
\end{equation}
where the $i$th qubit terms have $S_1$ symmetry, and all the coupling terms together form $D_n$ symmetry. Defining $H_{0}^{(i)} = B_z \cdot \frac{1}{2} \sigma_z^{(i)}$ and $H_{\text{cpl}} = c_{\text{cpl}}\cdot\frac{1}{4} \sum_{i=1}^{n}\sigma_z^{(i)}\sigma_z^{(i+1)}$, the discretized propagator at the $j$th time step in Eq.~\ref{exppropagator} can be Trotterized by the symmetry of the terms as
\begin{equation}
U_j = \prod_{i=1}^n \left[ \text{exp} \left(- i \tau \left( H_{0}^{(i)} + H_c^{\prime (i)}[(j + \frac{1}{2})\tau] \right) \right)\text{exp} \left(- i \frac{\tau}{n} H_{\text{cpl}} \right) \right] + O(n^2 \tau^2),
\label{propagator_cpl}
\end{equation}
where the first and second exponentials in the bracket have $S_1$ and $D_n$ symmetries, respectively. We transform each exponential term so that the number of blocks is maximized and the size of each time-dependent block is minimized. The transformation is given by
\begin{equation}
U_j \approx \prod_{i=1}^n \bigg[ A_i \text{exp} \left(- i \tau A_i^{\dagger} \left( H_{0}^{(i)} + H_c^{\prime (i)}[(j + \frac{1}{2})\tau] \right) A_i \right) A_i^{\dagger} A_D \text{exp} \left(- i \frac{\tau}{n} A_D^{\dagger} H_{\text{cpl}} A_D \right) A_D^{\dagger} \bigg],
\label{propagator_cpl_transformed}
\end{equation}
where $A_i = A_{(i, n)}$ is the permutation matrix that swaps the $i$th and the $n$th qubit. After transformation, $A_{(i, n)}^{\dagger} \sigma_\alpha^{(i)} A_{(i, n)}$ is equal to $\sigma_\alpha^{(n)}$ which has $2^{n-1}$ of $2 \times 2$ blocks that are exactly the same. The matrices $A_i$ and {$A_D \text{exp} (- i \frac{\tau}{n} A_D^{\dagger} H_{\text{cpl}} A_D ) A_D^{\dagger}$} are constant and only need to be calculated one time. As such, we reduce the time-dependent term from one $2^n \times 2^n$ matrix to $n$ of $2 \times 2$ matrices.

\begin{figure}[htbp]
\centering
    \begin{subfigure}{0.48\textwidth}
    \centering
    \includegraphics[width=1.0\textwidth]{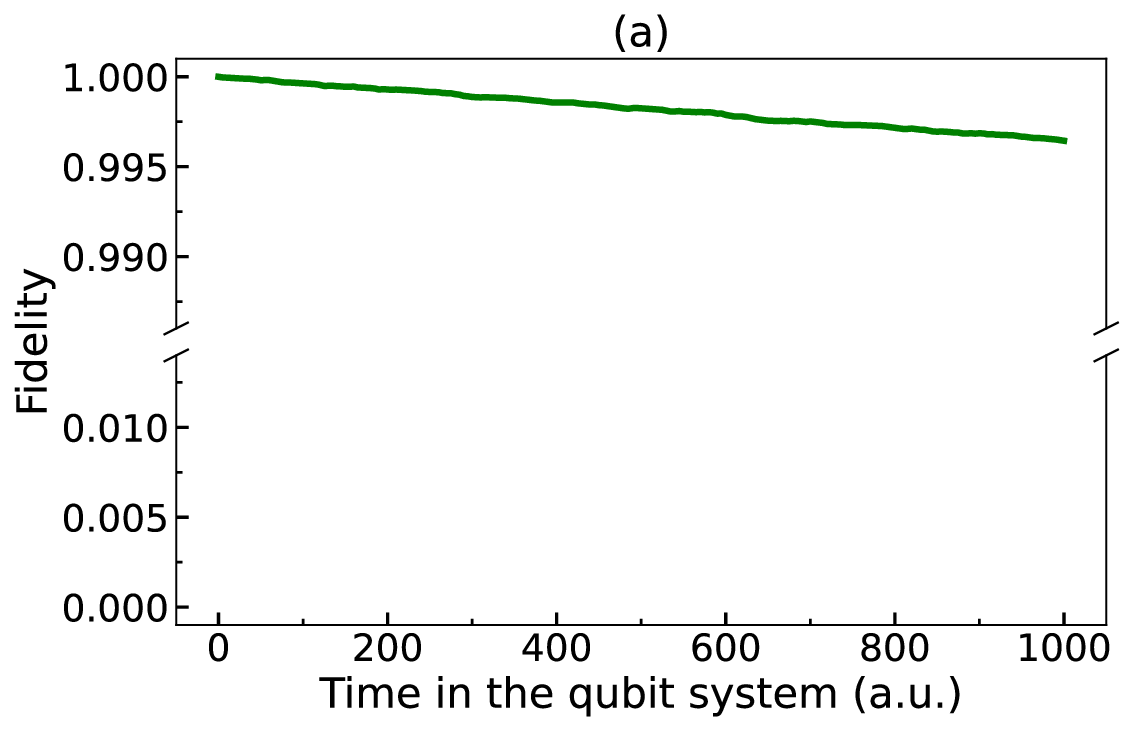}
    \end{subfigure}
    ~
    \begin{subfigure}{0.48\textwidth}
    \centering
    \includegraphics[width=1.0\textwidth]{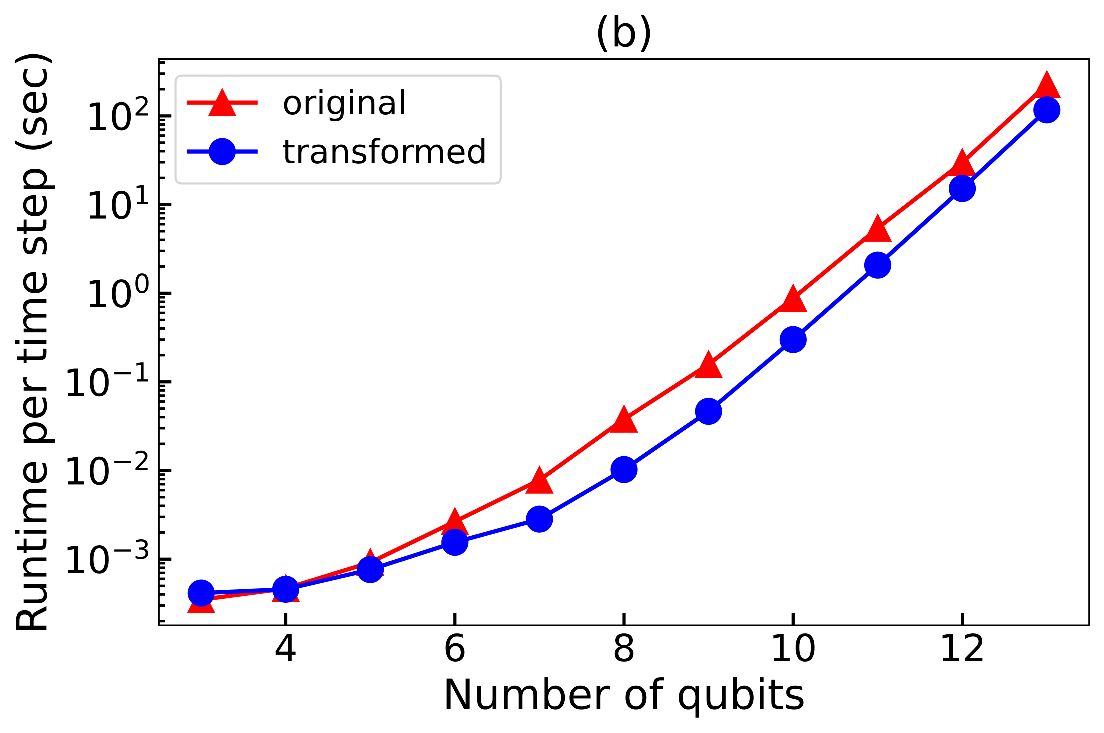}
    \end{subfigure}
\caption{\textbf{Fidelity and computational runtime of Trotterized propagator.} (a) Fidelity $F$ in the $11$-qubit system as a function of time. (b) Comparison of the computational runtime per time step for $n$ qubits ranging from 3 to 13.}
\label{fig_LTS_comparison}
\end{figure}

To demonstrate {that the Trotterized and transformed propagator in Eq.~\ref{propagator_cpl_transformed} is a good approximation to the original propagator in the same form as Eq.~\ref{exppropagator}}, we let an $n$-qubit ($3 \leq n \leq 13$) system evolve for $20,000$ time steps with $\tau = 0.05$ a.u. We then evaluate the fidelity $F = \vert \frac{\text{Tr}( {K_j^{\text{LTS}}}^{\dagger} K_j^{\text{ori}} )} {2^n}\vert^2$ \cite{Wang2022} of the unitary matrix $K_j^{\text{ori}} = \prod_{m=j}^1 U_m^{\text{ori}}$ calculated with the original propagator in Eq.~\ref{exppropagator} and the unitary matrix $K_j^{\text{LTS}} = \prod_{m=j}^1 U_m^{\text{LTS}}$ calculated with the Trotterized propagator in Eq.~\ref{propagator_cpl_transformed}. Fig.~\ref{fig_LTS_comparison}a shows that in the $11$-qubit system, the fidelity $F$ is always above $0.996$, which is still highly accurate. Tests with other numbers of qubits yield similar accuracy. We also compared the runtime for calculating the original and transformed propagator per time step. As Fig.~\ref{fig_LTS_comparison}b shows, the transformed propagator is more time-efficient since each exponential term in Eq.~\ref{propagator_cpl_transformed} is block diagonalized into exactly the same blocks and we need to calculate the matrix exponential of only one block. Collectively, the tests above show that the transformed propagator in Eq.~\ref{propagator_cpl_transformed} is highly accurate and time-efficient. In Sec.~\ref{general_framework_LTS_symmetry}, we introduce a general framework for parallel computing with Lie-Trotter-Suzuki decomposition and symmetry-based transformation that can be applied to nearly all Hamiltonians of multi-qubit systems.

\section{Conclusion} \label{conclusion}

In conclusion, we have harnessed the intrinsic symmetry of finite groups to accelerate quantum optimal control calculations in multi-qubit systems. The homogeneity and distinguishability of the qubits, resulting in the symmetry of multi-qubit systems, are ubiquitous in nearly all multi-qubit systems, which allows us to generalize our approach to a variety of quantum computing configurations. Our results show that even in the case of inseparable multi-qubit systems, it is possible to decompose the Hilbert space $\mathcal{H}(\mathbb{C}^{2^n})$ into a direct sum of orthogonal and complete subspaces. The selection rules intrinsic to the finite group symmetry restrict the transitions within each subspace. We also propose a scheme of quantum error suppression and quantum simulation in the symmetry-protected subspaces. In addition to these techniques, we developed a scheme to generalize our symmetry-based Hamiltonian transformation to general systems with the Lie-Trotter-Suzuki decomposition, which is naturally amenable to parallel computing. Taken together, our approach does not impose constraints to satisfy features of any specific quantum platform, which enables our symmetry-based approach to be easily used for general QOC calculations up to 14 qubits and beyond.

\section{Data Availability}
The code used for optimal control of multi-qubit systems is available at \url{https://github.com/xwang056/qoc_multi-qubits}.

\section{Supplementary Material}
See the supplementary material for additional mathematical details on dynamics of multi-qubit systems; gradient-based quantum optimal control algorithms; transformation of Hamiltonians with $S_n/D_n$ symmetry; mathematical proof of the orthogonality and completeness of the basis generated by the $D_n$-induced decomposition of the Hilbert space; ladder operators/selection rules.; additional sparsity plots for Hamiltonians; plots of optimal control pulses; power spectra/convergence plots of other qubit systems.

\begin{acknowledgments}
This work was supported by the U.S. Department of Energy, National Energy Technology Laboratory (NETL) under Award No. DE-FE0031896.
\end{acknowledgments}

\section{Competing Interests}
The Authors declare no competing financial or non-financial Interests.

\section{Author Contributions}
X.W.: conceptualization, methodology, software, validation, formal analysis, investigation, data curation, writing (original draft preparation), writing (review and editing), visualization; M.S.O.: methodology, formal analysis, investigation, writing (original draft preparation), writing (review and editing), supervision; A.K.: methodology, formal analysis, investigation, writing (original draft preparation); B.M.W.: conceptualization, methodology, formal analysis, investigation, writing (original draft preparation), writing (review and editing), project administration, funding acquisition.

\appendix

\section{Generation of the Adjoint Matrices in the Symmetry-Based Transformation} \label{generation_A}

We briefly present our procedure for generating the adjoint matrix $A_S$ or $A_D$ that block diagonalizes the Hamiltonians of an $n$-qubit system with $S_n$ or $D_n$ symmetry. The columns in the adjoint matrix are the orthonormal basis of the subspaces after the $S_n$- or $D_n$-induced decomposition.

When the coupling coefficient $c_{\text{cpl}}$ is zero, the angular momentum $J$ and its projection onto the $z$-axis $M$ are good quantum numbers. Therefore, the orthogonal basis $\vert J, M\rangle$ can be generated in each subspace with the Clebsch-Gordan coefficients of SU(2) \cite{ma2007, han1987, li2019}. An alternative way to generate the orthogonal basis is to use the irreducible representations (irreps) of $S_n$. Each irrep, denoted as $A^{\lambda}$, can be characterized with a standard Young diagram $\lambda$. The key procedure is to generate the operator $O^{\lambda}_j$ in the group algebra $\mathcal{R}_{S_n}$ for each unitary irrep $A^{\lambda}$ with the Young method as follows
\begin{equation}
O^{\lambda}_j=\sum_{i=1}^{n!}A^{\lambda}_{jj}(e_i)e_i, \ 1\leq j\leq d_{\lambda},
\label{groupAlgebraElementsSn}
\end{equation}
 where $A^{\lambda}_{jj}(e_i)$ is the $j$th diagonal element in the representation $A^{\lambda}(e_i)$ of the group element $e_i$, and $d_{\lambda}$ is the dimension of the irrep $A^{\lambda}$. Acting $O^{\lambda}_j$ on proper Fock states of the $n$-qubit system, $\{\vert\uparrow\rangle,\vert\downarrow\rangle\}^{\otimes n}$, the complete set of the orthogonal basis of each subspace, denoted as $\mathcal{H}_j^{\lambda}$, can be generated. The dimension of each subspace is $O(n)$. Using the Clebsch-Gordan coefficients of SU(2) and the Young method are mathematically equivalent and generate the same orthogonal basis in each subspace (this equivalence arises because the Young method was developed to generate the irreps of special unitary groups). \cite{ma2007, han1987, li2019} Additional details of the Young method and the Clebsch-Gordan coefficients of SU(2) can be found in Secs.~IB and IC in the Supplementary Material.

When the coupling coefficient $c_{\text{cpl}}$ is nonzero, the symmetry of the multi-qubit system reduces to $D_n$. We denote each irrep of $D_n$ as $A^{\theta}$ and its dimension as $d_{\theta}$. In this case, we can define the following operator
\begin{equation}
O^{\theta}_j=\sum_{i=1}^{2n}A^{\theta}_{jj}(e_i)e_i, \ 1\leq j\leq d_{\theta}
\label{groupAlgebraElementsDn}
\end{equation}
in the group algebra $\mathcal{R}_{D_n}$ for each unitary irrep $A^{\theta}$. Similarly, acting $O^{\theta}_j$ on proper Fock states of the $n$-qubit system, the complete set of the orthogonal basis of each subspace, denoted as $\mathcal{H}_j^{\theta}$, can be generated. The dimension of each subspace is $O(\frac{2^n}{n})$. Additional details of the $D_n$-symmetry-based method can be found in Secs.~ID and IE in the Supplementary Material. {Similar to the $D_n$-induced decomposition of the Hilbert space, multi-qubit systems having the symmetry of other finite groups than $S_n$ and $D_n$ can also be analyzed and simplified with the operators in the corresponding group algebra.}

The decomposition of the Hilbert space, i.e., $\mathcal{H} (\mathbb{C}^{2^n}) = \bigoplus_{\lambda, j} \mathcal{H}_j^{\lambda}$ or $\mathcal{H} (\mathbb{C}^{2^n}) = \bigoplus_{\theta, j} \mathcal{H}_j^{\theta}$, makes it possible to generate the adjoint matrix that transforms the Hamiltonians to block diagonal matrices. In the main text, we denote these two decompositions as $\mathcal{H} (\mathbb{C}^{2^n}) = \bigoplus_k \mathcal{H}_k^{S}$ and $\mathcal{H} (\mathbb{C}^{2^n}) = \bigoplus_k \mathcal{H}_k^{D}$ for conciseness.

\section{General Framework for Parallel Computing with Trotterization and the Symmetry-Based Transformation} \label{general_framework_LTS_symmetry}

We introduce a general framework for parallel computing with the Lie-Trotter-Suzuki decomposition and the symmetry-based transformation of the Hamiltonian of any multi-qubit system. The propagator of the quantum system is Trotterized so that terms sharing the same symmetry are put together and block diagonalized by the same adjoint matrix. The principle of block diagonalizing each exponential term is to maximize the number of blocks and minimize the size of each time-dependent block. Typically, the blocks are repetitive if the number of interacting qubits in the exponential term is smaller than $n$.

We present several examples to further illustrate the method above. The transformation of terms of the form $\sigma_\alpha^{(i)}, \alpha = x, y, z$ is shown in Sec.~\ref{LTS_decomposition}. The coupling terms having the form of $\sigma_\alpha^{(i)} \sigma_\alpha^{(j)}, i \neq j$ can be transformed with $A_{(i, n-1)} A_{(j, n)}$. The transformed term $A_{(j, n)}^{\dagger} A_{(i, n-1)}^{\dagger} \sigma_\alpha^{(i)} \sigma_\alpha^{(j)} A_{(i, n-1)} A_{(j, n)}$ is equal to $\sigma_\alpha^{(n-1)} \sigma_\alpha^{(n)}$ and has $2^{n-2}$ blocks, with each block having a size of $4 \times 4$. We can block diagonalize this term further with its $S_2$ symmetry. More specifically, letting $A = A_{(i, n-1)} A_{(j, n)} (\mathbb{I}_{2^{n-2}} \otimes A_{S_2})$, each {$4 \times 4$} block in $A^{\dagger} \sigma_\alpha^{(i)} \sigma_\alpha^{(j)} A$ can be transformed into {one $1 \times 1$} and {one $3 \times 3$} block. Similarly, the {terms having the form of $\sigma_\alpha^{\otimes i-1} \otimes \sigma_\beta \otimes \sigma_\alpha^{\otimes j-i-1} \otimes \sigma_\beta \otimes \sigma_\alpha^{\otimes n-j}$} can be transformed with index permutation and $S_2 \otimes S_{n-2}$ symmetry. Note that $\mathbb{I}_2, \sigma_x, \sigma_y, \sigma_z$ and their tensor products form the orthogonal basis of any {$2^n \times 2^n$} Hermitian matrix under the Hilbert-Schmidt inner product \cite{Cheverry2021}. Accordingly, the Hamiltonian of an $n$-qubit system can always be decomposed such that each component can be transformed into $\mathbb{I}_2^{\otimes l} \otimes \sigma_x^{\otimes m} \otimes \sigma_y^{\otimes p} \otimes \sigma_z^{\otimes q}, l + m + p + q = n$ by an adjoint matrix $A_I$, which permutes the indices. We can then transform this term with the adjoint matrix $A_G = \mathbb{I}_{2^{l}} \otimes A_{S_m} \otimes A_{S_p} \otimes A_{S_q}$ where $G = S_m \otimes S_p \otimes S_q$ is the finite group indicating the symmetry of this term. In summary,  a general Hamiltonian $H$ can be decomposed into a sum $H = \sum_{G, I} ( H_{0}^{(G, I) } + H_c^{(G, I) } )$ by the symmetry characterized by the finite group $G$ and the indices of the qubits $I$ ($I$ denotes qubits coupled to either static fields or controlling pulses simultaneously). The propagator of $H_{0}^{(G, I) } + H_c^{(G, I) }$ can be transformed by the adjoint matrix $A_{G, I} = A_I A_G$. As such, a general symmetry-based transformed and Trotterized propagator can be written as
\begin{align}
\begin{aligned}
U_j &= \prod_{G, I} \left[ \text{exp} \left(- i \tau \left( H_{0}^{(G, I) } + H_c^{(G, I) }[(j + \frac{1}{2})\tau] \right) \right) + O(\tau^2) \right] \\
&\approx \prod_{G, I} \left[ A_{G, I} \text{exp} \left(- i \tau A_{G, I}^{\dagger} \left( H_{0}^{(G, I) } + H_c^{(G, I) }[(j + \frac{1}{2})\tau] \right) A_{G, I} \right) A_{G, I}^{\dagger} \right].
\label{propagator_general}
\end{aligned}
\end{align}
When $A_G = \mathbb{I}_{2^{l}} \otimes A_{G_{n-l}}$ satisfies $l \geq 1$, the blocks repeat themselves $2^l$ times in the transformed Hamiltonian, allowing us to calculate the exponential of a {$2^{n-l} \times 2^{n-l}$} matrix rather than that of a full {$2^n \times 2^n$} matrix when computing the propagator.

It is worth noting that each exponential in the Trotterized propagator in Eq.~\ref{propagator_general} is independent of the others, which allows them to be trivially computed in parallel. Also, the exponential of each block (not counting the repetitive blocks) is independent, allowing us to parallelize the computation further. In Sec.~\ref{LTS_decomposition}, when testing the runtime for computing Eq.~\ref{propagator_cpl_transformed}, we calculated the exponentials in the transformed propagator in series. To approximate the runtime in an in-parallel computing setup, we divided the runtime for calculating the transformed propagator by $n$.

All of the $A_{G, I}$ adjoint matrices and the {blocks} in all of the exponentials are unitary, allowing us to easily calculate the transformed propagator's inverse in parallel. Also, the derivative of each exponential $\text{exp} (- i \tau A_{G, I}^{\dagger} ( H_{0}^{(G, I) } + H_c^{(G, I) }[(j + \frac{1}{2})\tau] ) A_{G, I} )$ with respect to the time-dependent control $B_\alpha[(j + \frac{1}{2})\tau]$ can be approximated as
\begin{align}
\begin{aligned}
&\frac{\text{d} \left( \text{exp} \left(- i \tau A_{G, I}^{\dagger} \left( H_{0}^{(G, I) } + H_c^{(G, I) }[(j + \frac{1}{2})\tau] \right) A_{G, I} \right) \right)}{\text{d} (B_\alpha[(j + \frac{1}{2})\tau] )} \\
\approx &-i \tau A_{G, I}^{\dagger} \tilde{H}_c^{(G, I) } A_{G, I} \cdot \text{exp} \left(- i \tau A_{G, I}^{\dagger} \left( H_{0}^{(G, I) } + H_c^{(G, I) }[(j + \frac{1}{2})\tau] \right) A_{G, I} \right)
\label{propagator_differentiation}
\end{aligned}
\end{align}
when $\tau$ is small and the control Hamiltonian has the simple expression of $H_c^{(G, I) }[(j + \frac{1}{2})\tau] = B_\alpha[(j + \frac{1}{2})\tau] \cdot \tilde{H}_c^{(G, I) }$, which is a common situation. As such, we can easily apply the transformed propagator $U_j$ in Eq.~\ref{propagator_general} to gradient-based methods with backpropagation.

\bibliographystyle{unsrt}
\bibliography{qubits}

\end{document}